\documentclass[11pt]{article}
\usepackage[british]{babel}
\usepackage[utf8]{inputenc}

\usepackage{amsmath, amsthm, amssymb,slashed}
\usepackage{amsfonts, amscd,latexsym,amsbsy,bm}
\usepackage[normalem]{ulem}
\usepackage{mathtools}
\usepackage{stackengine}
\usepackage{stmaryrd}
\usepackage{mathrsfs}
\numberwithin{equation}{section}

\usepackage{dsfont}
\usepackage{upgreek}

\usepackage{dutchcal}

\usepackage{graphicx,shortvrb}
\usepackage{xcolor}
\usepackage{tikz,pgfplots}
\usetikzlibrary{calc,arrows,arrows.meta,cd,shapes.geometric,positioning,through,decorations.markings,decorations.text}
\tikzset{>=latex}

\usepackage{cite}
\usepackage[colorlinks=true,linkcolor=blue, linktoc=page, urlcolor=blue,citecolor=magenta,pagebackref=true]{hyperref}
\renewcommand*{\backref}[1]{}
\renewcommand*{\backrefalt}[4]{%
  \ifcase #1%
  \or [Page~#2.]%
  \else [Pages~#2.]%
  \fi%
}

\newcommand{\Bpl}{\Big{(}}
\newcommand{\Bpr}{\Big{)}}
\newcommand{\Bbpl}{\Bigg{(}}
\newcommand{\Bbpr}{\Bigg{)}}

\newcommand{\RR}{\mathbb{R}}

\newcommand{\GL}{\operatorname{GL}}
\newcommand{\Ort}{\operatorname{O}}

\newcommand{\dpb}[2]{[\![#1\,,#2]\!]}

\def\be{\begin{equation}}
\def\ee{\end{equation}}

\def\calb{{\cal B}}

\def\calg{{\cal G}}
\def\calh{{\cal H}}

\def\calk{{\cal K}}
\def\call{{\cal L}}
\def\calm{{\cal M}}

\def\cals{{\cal S}}

\def\a{\alpha}
\def\b{\beta}

\def\l{\lambda}

\def\k{\kappa}

\def\m{\mu}
\def\n{\nu}

\def\r{\rho}

\def\s{\sigma}

\def\hc{{h}}
\def\pc{{p}}

\def\cJ{{J}}
\def\cH{{H}}
\def\cB{{B}}
\def\cP{{P}}

\begin{document}

\begin{titlepage}
 \vskip 1.8 cm

\begin{center}{\huge \bf Particle dynamics on torsional galilean spacetimes}\\

\end{center}
\vskip 1.5cm

\centerline{\large {{\bf José Figueroa-O'Farrill$^{1}$, Can Görmez$^{2}$ and  Dieter Van den Bleeken$^{2,3}$}}}

\vskip 1.0cm

\begin{center} 

    1) School of Mathematics and Maxwell Institute\\
    The University of Edinburgh\\
    Edinburgh EH9 3FD, Scotland, UK

    \vskip 0.5cm
    
    2) Primary address:\\
    Physics Department, Boğaziçi University\\
    34342 Bebek / Istanbul, Turkey
    
    \vskip 0.5cm
    
    3) Secondary address:\\
    Institute for Theoretical Physics, KU Leuven\\
    3001 Leuven, Belgium
    
    \vskip 0.5cm
    
    \texttt{j.m.figueroa@ed.ac.uk, can.gormez@boun.edu.tr, dieter.van@boun.edu.tr}

\vspace{0.5cm}

\end{center}
\vskip 0.5cm \centerline{\bf Abstract} \vskip 0.2cm \noindent We study
free particle motion on homogeneous kinematical spacetimes of galilean
type. The three well-known cases of Galilei and (A)dS--Galilei
spacetimes are included in our analysis, but our focus will be on the
previously unexplored torsional galilean spacetimes.  We show how in
well-chosen coordinates free particle motion becomes equivalent to
the dynamics of a damped harmonic oscillator, with the damping set
by the torsion. The realization of the kinematical symmetry algebra in
terms of conserved charges is subtle and comes with some interesting
surprises, such as a homothetic version of hamiltonian vector fields
and a corresponding generalization of the Poisson bracket.  We show
that the Bargmann extension is universal to all galilean kinematical
symmetries, but also that it is no longer central for nonzero
torsion.  We also present a geometric interpretation of this fact
through the Eisenhart lift of the dynamics.

\end{titlepage}
\tableofcontents

\section{Introduction}
\label{sec:introduction}

Colloquially and in a physics context, a kinematical Lie algebra is a
Lie algebra containing time and space translations, rotations and
boosts, with the assumption that boosts and space translations are
vectors under rotations, while time translations are invariant under
rotations.  Examples of kinematical Lie algebras are the isometry Lie
algebras of the maximally symmetric lorentzian manifolds (Minkowski
and (anti)de~Sitter spacetimes), the Galilei and Carroll algebras, the
Newton--Hooke algebras, as well as a host of other anonymous Lie
algebras which have been known since the pioneering work of Bacry and
Lévy-Leblond \cite{MR0238545,MR857383}.

The Lie groups corresponding to these kinematical Lie algebras
act transitively on so-called kinematical spacetimes, which have
recently been classified \cite{Figueroa-OFarrill:2018ilb}.  Among
them we find, of course, Minkowski and  (anti) de~Sitter spacetimes, but
also their nonrelativistic limits: Galilei and (anti)
de~Sitter--Galilei spacetimes, as well as their ultra-relativistic
limits: Carroll and (anti) de~Sitter--Carroll spacetimes.  There are
many more, but they all fall into families: lorentzian, galilean,
carrollian and aristotelian, depending on the invariant geometric
structure that they possess.

It is one of the postulates of general relativity that free particle
motion in a lorentzian spacetime (such as Minkowski or (anti)
de~Sitter) follows (causal) geodesics of the Levi-Civita connection,
and it is therefore a natural question to ask about free particle
motion in the other non-lorentzian kinematical spacetimes.

For example, the dynamics of a standard nonrelativistic free
particle, with Lagrangian $L=\frac{1}{2}\dot x^a \dot x^a$, is
well-known to be invariant under the Galilei group.  This can be
explained by the fact that the extrema of that Lagrangian can be
interpreted as geodesics of the invariant connection on Galilei
spacetime which is compatible with the invariant galilean
structure. Since this spacetime is a homogeneous spacetime for the
Galilei group there is a natural action of this group and because the
connection is invariant, the group acts as symmetries of geodesic
motion.

With one exception --- namely, the carrollian lightcone --- all
kinematical spacetimes have invariant connections
\cite{Figueroa-OFarrill:2018ilb,Figueroa-OFarrill:2019sex} and hence
we may define free particle motion on a kinematical spacetime as
geodesics relative to any invariant compatible connection.

In this paper we shall be concerned only with galilean kinematical
spacetimes.  In general spacetime dimension (here $\geq 4$), there are
three symmetric spatially isotropic homogeneous galilean spacetimes:
Galilei and (anti) de~Sitter--Galilei and two one-parameter families
of torsional galilean spacetimes with a common galilean limit.  As
mentioned above, free particle motion on Galilei spacetime is
understood.  Similarly, one can study free particle motion on (anti)
de~Sitter--Galilei spacetime, either intrinsically or by taking the
nonrelativistic limit of geodesic motion on (anti) de~Sitter
spacetime.  These were studied in \cite{Gibbons:2003rv}, who called
them Newton--Hooke spacetimes, and showed that the effect of the
cosmological constant $\Lambda$ was to modify the standard
nonrelativistic free particle Lagrangian by the addition of a
potential $L = \frac12 \dot x^a \dot x^a + \frac12 \Lambda x^a x^a$,
such that force is restorative for $\Lambda < 0$ (anti~de~Sitter) and
repulsive for $\Lambda > 0$ (de~Sitter).  Motion on (anti)
de~Sitter--Galilei spacetime has recently been considered in
\cite{Maxfield:2022hkd} in the context of AdS/CFT, see also \cite{Bizon:2018frv} for an earlier discussion of the invariant wave equation.

Particle motion in the torsional galilean spaces had remained
unexplored until now and this was the main motivation for this work.
As we will see, the effect of the torsion is simply to add a dampening
force to the motion. The damped harmonic oscillator has a long
history, reviewed, e.g., in \cite{MR645698}. Our novel perspective --
to interpret it as motion invariant under a torsional galilean
kinematical symmetry algebra -- reveals some relations among the
conserved charges that so far went unnoticed. Although the galilean
kinematical symmetries act on phase space as usual, i.e., as a
subgroup of diffeomorphisms, the time translation is neither
hamiltonian nor symplectic when the torsion is non-zero. Since it
remains homothetically hamiltonian one can still associate a conserved
charge, but the Poisson algebra of conserved charges is no longer
homomorphic to the algebra of vector fields generating the kinematical
symmetries. We show however that there exists a natural Lie bracket on
the space of phase space functions extended with a scaling weight,
that makes this space homomorphic as a Lie algebra to the algebra of
homothetic hamiltonian vector fields. Under this bracket the conserved
charges do form an algebra homomorphic to the torsional galilean
symmetry algebra. More precisely the algebra of conserved charges is a
one-dimensional extension, well known in the case without torsion as
the Bargmann extension \cite{bargmann1954unitary}, of the kinematical
algebra. When the torsion is non-zero this Bargmann extension is no
longer central. We proceed to show that, as in the case without
torsion \cite{Duval:1984cj}, this extension can be given a geometric
interpretation through the Eisenhart lift
\cite{Eisenhart1928_10.2307/1968307}.

The results in this paper lead to a few natural questions that could
be interesting to investigate further. How does the galilean
kinematical Lie algebra embed into the full symmetry group of the
damped harmonic oscillator \cite{cervero1984quantum}? What is the
r\^ole of the torsional kinematical symmetry and its (non-central)
Bargmann extension in the quantum theory? Is the notion
of free particle defined via geodesics of invariant connections in a
non-lorentzian geometry equivalent to the notion of elementary
particle in the sense of Souriau \cite{MR0260238,MR1461545}?

This paper is organized as follows. In Section~\ref{sec:kin-hom-sts}
we review the geometry of the torsional homogeneous galilean
spacetimes, deriving some coordinate expressions and paying particular
attention to the invariant connections. In particular, in
Section~\ref{sec:gal-klas} we define the spacetimes in terms of their
Klein pairs, in Section~\ref{sec:mod-exp} we introduce modified
exponential coordinates which we prove to be global coordinates and in
Section~\ref{sec:invcon} we give explicit expressions for the
Christoffel symbols of the invariant connections in these coordinates.
In Section~\ref{sec:dho} we discuss geodesic motion on the galilean
spacetimes relative to the invariant connection. More concretely, in
Section~\ref{sec:dho-from-geodesics} we show that in a convenient
parametrization, the geodesic equation relative to any invariant
connection reduces to that of a damped harmonic oscillator. Then in
Section~\ref{sec:charges}, we discuss the realization of the
kinematical Lie algebra as symmetries of this equation and in terms of
conserved charges. In Section~\ref{sec:NCbarg} we re-examine geodesic
motion in terms of motion in a Newton--Cartan geometry and we relate
it, via the Eisenhart lift, to null geodesic motion on certain
homogeneous pp-waves. The paper ends with four appendices. In
Appendix~\ref{sec:type1} we record some definitions about Type~I
Newton--Cartan geometry, particularly the notion of compatible
NC-doublets and NC-triplets and their use in defining an affine
connection compatible with the Newton--Cartan structure. In
Appendix~\ref{sec:homap} we discuss the notion of symplectic and
hamiltonian homotheties, which play an important rôle in
Section~\ref{sec:charges}. In particular we motivate and define a
modified bracket on phase space functions, which we believe is new and
might find applications outside the context of this paper as well.
In Appendix~\ref{sec:lowdap} we extend the discussion of the
main text to galilean spacetimes in spacetime dimension $\leq 3$,
where particularly in dimension $3$ there is a richer family of
homogeneous galilean spacetimes. Finally, in Appendix \ref{confapp} we comment on the freedom to conformally redefine the Eisenhart lift and use this to connect some of our results to those of \cite{Figueroa-OFarrill:2022ryd}.

\section{Kinematical homogeneous spacetimes of galilean type}
\label{sec:kin-hom-sts}

In this section we review the kinematical algebras of galilean type,
the associated homogeneous spacetimes and the key invariant geometric
structure -- the invariant NC-compatible connection -- that
differentiates them. With the aim of keeping the discussion brief and
as concrete as possible, we only summarize those results of relevance
to the remainder of the paper and do so in a language and notation
adapted to the problem discussed there. We refer to
\cite{Figueroa-OFarrill:2018ilb,Figueroa-OFarrill:2019sex} for a more
general discussion of kinematical homogeneous spaces of all types, as
well as a more precise and detailed discussion of the galilean case.

\subsection{Kinematical Lie algebras of galilean type}
\label{sec:gal-klas}

Our starting point is the Lie algebra of (infinitesimal) symmetries in
$d$ spatial dimensions, a generalization of the well-known galilean
algebra of a nonrelativistic free particle.

A \emph{kinematical Lie algebra of galilean type} is a real Lie
algebra for which there exists a basis  $\{J_{ab}=-J_{ba}, B_a, H,
P_a\}$, $a,b=1,\ldots, d$ with Lie brackets
\begin{equation}\label{isotropybrackets}
  \begin{split}
    [J_{ab},J_{cd}] &= \delta_{bc} J_{ad}-\delta_{ac}J_{bd}-\delta_{bd}J_{ac}+\delta_{ad}J_{bc}\\
    [J_{ab},B_c] &= \delta_{bc}B_a-\delta_{ac}B_b\\
    [J_{ab},P_c] &= \delta_{bc}P_a-\delta_{ac}P_b\\
    [J_{ab},H] &= 0
\end{split}
\end{equation}
and\footnote{Our definition here and our further discussion in the
  main text is only complete when $d\geq 3$. When $d=2$ there exists
  an additional kinematical algebra of galilean type where the
  brackets \eqref{galbrackets} are slightly modified. See
  Appendix~\ref{sec:lowdap} for the discussion when $d\leq 2$.}
\begin{equation}\label{galbrackets}
  [H,B_a]=-P_a\qquad [H,P_a]=\alpha B_a+\beta P_a\qquad [P_a,B_b]=[P_a,P_b]=[B_a,B_b]=0
\end{equation}
where $\alpha,\beta$ are arbitrary real constants. We refer to this
Lie algebra as $\mathfrak{k_{(\a,\b)}}$.

The first set of brackets~\eqref{isotropybrackets} is universal to all
kinematical lie algebras, by their very definition. The $J_{ab}$
generate a rotation subalgebra under which the generators $B_a$ and
$P_a$ transform as vectors, while $H$ is a scalar.  A basis
independent definition of a \emph{kinematical Lie algebra} is a Lie
algebra that contains an $\mathfrak{so}(d)$ subalgebra with respect to
which the whole algebra decomposes as $\mathfrak{so}(d) \oplus 2V
\oplus  S$ where $2V$ are two copies of the $d$-dimensional (vector)
irreducible representation of $\mathfrak{so}(d)$ and $S$ is the
one-dimensional (scalar) trivial representation of $\mathfrak{so}(d)$.
Some well known examples other than the galilean ones we restrict
attention to in this paper, are the Poincaré and Carroll algebras.

The Lie brackets between the generators $\{H, B_a, P_a\}$ are left
free by the definition of kinematical Lie algebra and are only
constrained by the Jacobi identity. The brackets \eqref{galbrackets},
which are only a subset of the possible solutions to the Jacobi
identities, then select those kinematical algebras that we call `of
galilean type'.\footnote{For a classification of (spatially isotropic)
  kinematical Lie algebras, see
  \cite{MR0238545,MR857383,Figueroa-OFarrill:2017ycu} (for $d=3$),
  \cite{Figueroa-OFarrill:2017tcy} (for $d>3$) and
  \cite{Andrzejewski:2018gmz} (for $d=2$).  For $d=1$ every
  three-dimensional Lie algebra is kinematical, so the classification
  goes back to Bianchi \cite{Bianchi,MR1900159}.  Note that in the
  literature the `type' nomenclature is often reserved for Klein pairs
  $(\mathfrak{k},\mathfrak{h})$ of a kinematical Lie algebra
  $\mathfrak{k}$ together with a subalgebra $\mathfrak{h}$, see, e.g.,
  \cite{Figueroa-OFarrill:2018ilb}, where Klein pairs are referred to
  as Lie pairs. These pairs fall into the distinct
  classes of lorentzian-, riemannian-, galilean-, carrollian-,
  aristotelian- and low-dimensional exotic- type. One can extend this
  typification of the pairs to the kinematical algebras themselves, by
  defining such an algebra to be of a particular type if it allows a
  Klein pair of that type. The only subtlety is that with this
  definition a kinematical Lie algebra can be of more than one type. For
  example, the Poincaré algebra is of both lorentzian and carrollian
  type, as illustrated in the example on page 9 of
  \cite{Figueroa-OFarrill:2018ilb} or more recently also in
  \cite{Figueroa-OFarrill:2021sxz}, which displays a number of
  homogeneous spaces of the Poincaré group of different types
  describing the asymptotic geometry of Minkowski spacetime.
  Fortunately, there is no such ambiguity for the galilean type: only
  galilean Lie algebras admit (spatially isotropic) galilean Klein
  pairs.} Indeed, the class of algebras $\mathfrak{k}_{(\a,\b)}$
defined by \eqref{isotropybrackets} and \eqref{galbrackets}, contains
the Galilei algebra $\mathfrak{k}_{(0,0)}$ as well as the galilean
(A)dS -- or Newton--Hooke -- algebras $\mathfrak{k}_{(\pm 1, 0)}$.
Less familiar algebras can be obtained by choosing $\beta\neq 0$ and
these will be the main focus of this paper. There is some redundancy
in the parameters $(\alpha, \beta)$, as well as through mixing of the
$P_a$ and $B_a$: different choices can lead to isomorphic Lie
algebras.  We will discuss this in more detail in the next subsection
at the level of the associated homogeneous spacetimes.

Finally let us mention that based on the above, one can define a
kinematical Lie group of galilean type simply to be a Lie group whose
Lie algebra is kinematical of galilean type. We'll refer to the
(simply connected) Lie group with Lie algebra $\mathfrak{k}_{(\a,\b)}$
as $\calk_{(\a,\b)}$.

\subsection{Homogeneous spacetimes and modified exponential coordinates}
\label{sec:mod-exp}

As a physicist, one would intuitively interpret the generators of the
kinematical algebra --- \eqref{isotropybrackets} and
\eqref{galbrackets} --- as spatial rotations $J_{ab}$, time and space
translations $H$, $P_a$ as well as boosts $B_a$. One should be aware
however that such interpretation is associated to an action of the
associated transformations on a spacetime. It is only this action
which distinguishes the boosts $B_a$ from the translations $P_a$:
while boosts leave the origin invariant the translations do not. If we
assume the group action to be transitive then the spacetime will be a
homogeneous space that can be identified with a coset space
$\calk/\calh$ of the kinematical Lie group $\calk$.

The mathematical formulation of our physical intuition is then that a
(homogeneous) kinematical spacetime corresponds\footnote{More formally
  a \emph{homogeneous kinematical spacetime} is a connected smooth
  manifold $\calm$ with a transitive (and locally effective) action by
  a kinematical Lie group $\calk$, whose typical stabilizer subgroup
  $\calh$ has a Lie algebra $\mathfrak{h}$ which is given by
  $\mathfrak{so}(d) \oplus V$ as an $\mathfrak{so}(d)$ representation. It follows from this definition
  that any homogeneous kinematical space-time $\calm$ is
  ($\calk$-equivariantly) diffeomorphic to the coset space
  $\calk/\calh$. This then allows to identify it with the pair
  $(\mathfrak{k}$, $\mathfrak{h})$ given a number of further technical
  conditions, such as the pair  being \emph{effective} and
  \emph{geometrically realizable}. Full details can be found in e.g. 
  \cite{Figueroa-OFarrill:2018ilb}.} to a \emph{Klein pair}
$(\mathfrak{k}, {\mathfrak{h}})$. Such that $\mathfrak{k},
{\mathfrak{h}}$ are the Lie algebras of $\calk$ and its subgroup
$\calh$ respectively, and we require $\mathfrak{h}$ to contain
precisely the rotations $\mathfrak{so}(d)$ and one vector
representation $V$, i.e.
\begin{equation}
  \mathfrak{h}=\mathfrak{so}(d)\oplus V \subset \mathfrak{k}=\mathfrak{so}(d)\oplus 2V\oplus S
\end{equation}
This last condition imposes that rotations and one of its vector
representations, the boosts, leave spacetime points invariant (i.e.
generate the stabilizer subgroup). The complement $V\oplus S$ are the
space translations (a vector) and the time translation (a scalar). It
is the choice of $\mathfrak{h}$ which distinguishes the boosts from
translations inside the subspace $2V$ of $\mathfrak{k}$.

Since in this paper we will always have a homogeneous kinematical
spacetime -- i.e. a pair $(\mathfrak{k}, \mathfrak{h})$ -- in mind,
we'll simply indicate the choice of $\mathfrak{h}$ through our
notation for the basis $\{J_{ab},B_a,P_a,H\}$ of $\mathfrak{k}$. From
now on we'll assume that this notation implies that $\{J_{ab},B_a\}$
is a basis of $\mathfrak{h}$. In other words, with this additional
interpretation of the notation, the choice of basis
(\ref{isotropybrackets}, \ref{galbrackets}) for a kinematical Lie
algebra of galilean type directly defines a kinematical homogeneous
spacetime of galilean type. Since $(\alpha,\beta)$ are the only free
parameters in (\ref{isotropybrackets}, \ref{galbrackets}), which
furthermore do not appear in the brackets of $\mathfrak{h}$, we can
indicate the corresponding homogeneous spacetime as
\begin{equation}
\calm_{(\a,\beta)}=\calk_{(\a,\b)}/\calh\,.
\end{equation}

Let us now define global coordinates on $\calm_{(\a,\beta)}$, that we will refer to as {\it modified exponential coordinates}.  We may
choose the coset of the identity element $o=e\calh \in
\calm_{(\a,\beta)}$ as our origin. A point $p\in \calm_{(\a,\beta)}$
is then given the coordinates $(t,x^a)$ via the definition
\begin{equation}\label{modexp}
  p= e^{tH} e^{x^a P_a} \cdot o.
\end{equation}
This defines a smooth map $j: \RR^{d+1} \to \calm_{(\a,\beta)}$
sending $(t,x^a) \mapsto e^{tH} e^{x^a P_a} \cdot o$.  This map is a
local diffeomorphism.  Indeed, we can see that its derivative has full
rank by simply computing the pull-back of the left-invariant
Maurer--Cartan one-form on $\calk_{(\a,\b)}$ via the map $\sigma:
\RR^{d+1} \to \calk_{(\a,\b)}$ sending $(t,x^a) \mapsto e^{t H} e^{x^a P_a}$.
Doing so we find that
\begin{equation}
  \sigma^{-1} d\sigma = H dt + (dx^a + \beta x^a dt) P_a + \alpha x^a
  dt B_a,
\end{equation}
from where we read off the coframe $\theta = (dt, dx^a + \beta x^a
dt)$, which is clearly everywhere invertible.  Next we observe that
$\calm_{(\a,\b)}$ is acted on transitively by the solvable subgroup
$\calb_{(\a,\b)} \subset \calk_{(\a,\b)}$ generated by $H, B_a, P_a$.
In other words, the rotations are redundant and we will ignore them
for the purposes of showing that the coordinates $(t,x^a)$ are indeed
global.

We now define a transitive action of $\calb_{(\a,\b)}$ on $\RR^{d+1}$
by demanding that the map $j$ be equivariant.  The action of $\calb_{(\a,\b)}$ on $\calm_{(\a,\b)}$ is induced by left multiplication on the group, so calculating the product of exponentials we arrive at:
\begin{equation}\label{eq:group-action}
  \begin{split}
    e^{aH} \cdot (t,x^a) &= (t+a, x^a)\\
    e^{v^a B_a} \cdot (t,x^a) &= \left(t, x^a + e^{-t\beta/2} \frac{\sin(\omega t)}{\omega} v^a\right)\\
    e^{y^a P_a} \cdot (t, x^a) &= \left(t, x^a + e^{-t\beta/2} \left(\cos(\omega t) - \beta \frac{\sin(\omega t)}{2\omega}\right) y^a\right),
  \end{split}
\end{equation}
where $\omega := \tfrac12 \sqrt{4\alpha - \beta^2}$. This action is
transitive almost by definition. Indeed, starting from the origin
$(0,\mathbf{0})$ we can reach $(t,x^a)$ by acting with
$\sigma(t,x^a) = e^{t H} e^{x^a P_a}$. By equivariance, the image of
the map $j$ is an orbit of $\calb_{(\a,\b)}$ on $\calm_{(\a,\b)}$, but
since $\calm_{(\a,\b)}$ is homogeneous, $j$ is surjective and, being a
local diffeomorphism, it is a covering map.\footnote{Strictly speaking
  it is a branched covering, but this cannot happen for equivariant
  maps between homogeneous spaces, since the existence of a non-empty
  branched locus would spoil homogeneity.}  Since both $\RR^{d+1}$ and
$\calm_{(\a,\b)}$ are simply connected, the map $j$ must be a
diffeomorphism.

In summary, the coordinates $(t,x^a)$ thus defined are global and
hence, as manifolds, all the spacetimes $\calm_{(\a,\b)}$ are
diffeomorphic to $\mathbb{R}^{d+1}$.  Either directly from equation~\eqref{eq:group-action} or starting with \eqref{modexp} via the method in \cite{Figueroa-OFarrill:2019sex}, one calculates that the kinematical Lie algebra
$\mathfrak{k}_{(\a,\b)}$ is realized\footnote{Recall that in the
  standard conventions this is an \emph{anti}-homomorphism of Lie
  algebras.} on $\calm_{(\a,\b)}$ through the vector fields
\begin{equation}\label{kilvec}
  \begin{split}
    \xi_{J_{ab}} &= x^b\partial_a-x^a\partial_b\\
    \xi_{H} &= \partial_t \\
    \xi_{P_{a}} &= e^{-\beta t/2} \Bpl \cos \omega t - \frac{\beta}{2}\frac{\sin\omega t}{\omega} \Bpr \partial_a \\
  \xi_{B_{a}} &= e^{-\beta t/2} \frac{\sin\omega t}{\omega} \partial_a.
\end{split}
\end{equation}
Note that depending on the respective values of $\alpha, \beta$ the
parameter $\omega$ can be either real or imaginary, in all cases the
vector fields above remain real and well defined.

Remark that although the time translations are simply a shift of the
time coordinate $t$ by a constant, this is not the case for the
spatial translations which are a shift with a particular function of
time in case $\a$ or $\b$ are non-zero.

Although as manifolds all the kinematical homogeneous spacetimes of
galilean type $\calm_{(\a,\b)}$ are the same, this is no longer true if we equip them
with invariant geometric structures, since the symmetries act
differently in the different cases. We'll discuss this in more detail
in the next subsection.

Before we introduce these additional geometric invariants, let us
comment on the redundancy in the parameters $(\a,\b)$. Observe that
the change of basis $H\rightarrow s H, B_a \rightarrow s^{-1} B_a$,
for any $s\in\mathbb{R}^\times$, leads to the change of parameters
\begin{equation}
\alpha\rightarrow s^2\a\qquad \beta \rightarrow s \beta \label{abtransfo}
\end{equation}
This implies that the homogeneous spaces of which the parameters
$(\alpha, \beta)$ are related through such a transformation are
isomorphic\footnote{As homogeneous spaces, not just as manifolds.}.
Indeed this can be seen explicitly by observing that if one
accompanies the transformation \eqref{abtransfo} with the coordinate
transformation $(t,x^a)\rightarrow (s^{-1}t,x^a)$ then the vector
fields \eqref{kilvec} remain invariant.
  
Without loss of generality one can thus reduce the values of
$(\alpha,\beta)$ via \eqref{abtransfo} to one of the following three
classes
\begin{itemize}
\item Galilei spacetime: $\calm_{(0,0)}$
\item (torsional) galilean dS spacetime: $\calm_{(\gamma,1+\gamma)}\cong \calm_{(\gamma s^2,(1+\gamma)s)}\qquad \gamma\in[-1,1]$
\item (torsional) galilean AdS spacetime: $\calm_{(1+\chi^2,2\chi)}\cong \calm_{((1+\chi^2)s^2,2\chi s)}\quad \chi\in[0,\infty)$ 
\end{itemize}
The standard, rather well-known cases all have $\beta=0$ and are
Galilei spacetime ($\alpha=0$), galilean de Sitter ($\alpha<0$) and
galilean Anti de Sitter $(\a>0)$. The cases with $\beta\neq 0$ are
less well studied and are the topic of this paper. As we'll see in the
next subsection, $\beta$ parameterizes the torsion of the unique
invariant connection compatible with the invariant Newton--Cartan
structure on the $\calm_{(\a,\b)}$.

In the following we will find it convenient to keep working with the
parameters $(\a,\beta)$, rather than $\gamma$ or $\chi$ that
parametrize the inequivalent spacetimes, since it will allow us to
discuss all cases at once in a simple way. Furthermore each of $\a$
and $\beta$ will turn out to have a clear physical interpretation in
the particle dynamics, with $\alpha$ parameterizing a harmonic
potential and $\beta$ determining the damping.

\subsection{The invariant NC-compatible connection}
\label{sec:invcon}

The difference between the various kinematical homogeneous spacetimes
of galilean type $\calm_{(\a,\b)}$ is most explicitly seen in their
unique invariant NC-compatible connection. This is the unique
invariant affine connection that preserves the invariant Newton--Cartan
structure on $\calm_{(\a,\b)}$, as we'll explain below. In modified
exponential coordinates \eqref{modexp} the non-vanishing components of
the invariant NC-compatible connection on $\calm_{(\a,\b)}$ are
\begin{equation}\label{invcomp}
  \begin{split}
    \Gamma^a_{bt} &= \beta\delta^a_b  \\
    \Gamma^a_{tt} &= \alpha x^a.
  \end{split}
\end{equation}
Note that
since $\Gamma_{tb}^a=0$ this connection has torsion, the non-vanishing
components being \begin{equation}
  T_{bt}^a=\beta\delta^a_b
\end{equation}
Furthermore the only non-vanishing components of the Riemann tensor are
\begin{equation}
  R^a{}_{tbt}=-R^a{}_{ttb}=\alpha \delta^a_b
\end{equation}
We thus see that the parameters $(\a,\beta)$ can be identified with
respectively the curvature and torsion of the unique invariant
NC-compatible connection on $\calm_{(\a,\b)}$.

To derive the above results one starts from the fact
\cite{Figueroa-OFarrill:2019sex} that since the $\calm_{(\a,\b)}$ are
of galilean type, they carry an invariant Newton--Cartan structure\footnote{See appendix \ref{sec:type1} for a definition, nomenclature and some related notions.}
($\tau_\m, h^{\m\n}$).  (In fact, one can rescale $\tau$ and $h$
independently, by nonzero real numbers, so one has a two-parameter
family of invariant Newton--Cartan structures.)  In the
modified exponential coordinates $(x^\m)=(t,x^a)$ introduced in
\eqref{modexp}, the invariant Newton--Cartan structure takes the same
trivial form on all $\calm_{(\a,\b)}$:
\begin{equation}
\tau_\m=\delta_\m^t\qquad h^{\m\n}=\delta^\m_a\delta^\n_b \delta^{ab}\label{NCstructure}
\end{equation}
Crucially however the invariant affine connections differ
significantly in the various cases. Such connections can be classified
using the theorem of Nomizu \cite{nomizu}, or more explicitly by
demanding invariance under the infinitesimal coordinate
transformations associated to the kinematical symmetries\footnote{This
  amounts to solving $\mathcal{L}_\xi \nabla = 0$ for each $\xi$ in
  \eqref{kilvec}, which relative to local coordinates
  becomes the differential relations
  $\xi^\r\partial_\r\Gamma_{\m\n}^\l-\Gamma_{\m\n}^\r\partial_\r
  \xi^\l+\Gamma_{\r\n}^\l\partial_\m
  \xi^\r+\Gamma_{\m\r}^\l\partial_\n
  \xi^\r+\partial_\m\partial_\n\xi^\l=0$.}. Both methods agree,
and one finds that in modified exponential coordinates the only
non-zero components of an invariant affine connection
$\Gamma_{\m\n}^\r$ on $\calm_{(\a,\b)}$  are\footnote{The result
  \eqref{eq:invcon} is complete for $d\geq 3$.  Some
  exceptions are present when $d\leq 2$, see Appendix~\ref{sec:lowdap}.}
\begin{equation}\label{eq:invcon}
  \begin{split}
    \Gamma^t_{tt} &= (\kappa+\iota)  \\
    \Gamma^a_{tb} &= \kappa\delta^a_b \\
    \Gamma^a_{bt} &= (\beta + \iota)\delta^a_b  \\
    \Gamma^a_{tt} &= \alpha x^a.
  \end{split}
\end{equation}
In summary, on a given $\calm_{(\a,\b)}$ there is a family of
invariant connections parameterized by the two (unconstrained, real)
constants $\kappa, \iota$.

A short calculation reveals that among this two parameter family of
invariant connections there is a unique one that is compatible with
the invariant Newton--Cartan structure \eqref{NCstructure} on
$\calm_{(\a,\b)}$, i.e. such that $\nabla_\m\tau_\n=0$ and $\nabla_\m
h^{\n\r}=0$. This connection, that we refer to as the \emph{invariant
  NC-compatible connection}, is the invariant affine connection for
which $\k=\iota=0$, which leads to \eqref{invcomp}.

\section{Torsional galilean particles and the damped harmonic oscillator}
\label{sec:dho}

In this section we will provide a physical interpretation to the less
familiar galilean kinematical algebras with $\beta\neq 0$, by
realizing them as the symmetries of free, or geodesic, particle motion
in the space-time $\calm_{(\a,\b)}$. In the first part of this section
we show how this geodesic motion with respect to the invariant NC
compatible connection on $\calm_{(\a,\b)}$ can be identified, upon
fixing time parametrization invariance, with the dynamics of the
damped harmonic oscillator. In the second part of this section we
study how (an extension of) the kinematical symmetry algebra
$\mathfrak{k}_{(\a,\b)}$ is realized in terms of phase space vector
fields and conserved charges.

\subsection{Damped harmonic oscillator from geodesic equation}
\label{sec:dho-from-geodesics}

To formulate a $\calk_{(\a,\b)}$ invariant particle dynamics on
$\calm_{(\a,\b)}$ we can simply define the particle motion to be the
geodesic equation with respect to an invariant connection. Since such
invariant connections are not unique, see Section~\ref{sec:invcon},
there could a priori be more than one invariant particle dynamics. It
turns out however that the free parameters $\kappa, \iota$ specifying
the different invariant connections drop out of the actual geodesic
equations. In other words, all invariant connections share the same
set of geodesics, something which should not really be a surprise,
since in the highly symmetric situation we are considering one would
expect\footnote{Indeed this is the case.} the geodesics to be
generated purely by the symmetries, and these are independent of the
parameters $\kappa$ and $\iota$.

The affine geodesic -- or autoparallel -- equation associated to an
affine connection $\Gamma_{\m\n}^\r$ reads
\begin{align}
\frac{d^2 x^\rho}{d\s^2} + \Gamma^{\rho}_{\mu\nu} \frac{dx^\mu}{d\s} \frac{dx^\nu}{d\s} = f  \frac{dx^\rho}{d\s}\label{autoeq}
\end{align}
Choosing coordinates $(x^\m)=(t,x^a)$ we can re-express $f$ via the
$t$ component of the above equation:
\begin{equation}
 f= \frac{\frac{d^2t}{d\s^2} + \Gamma_{\mu\nu}^ t \frac{dx^\mu}{d\s} \frac{dx^\nu}{d\s} }{\frac{dt}{d\s}}
\end{equation}
Inserting this into \eqref{autoeq} and then specializing to the
invariant connections \eqref{eq:invcon} we
find that it is equivalent to 
\begin{equation}
\frac{d^2x^a}{d\s^2}+\beta \frac{dx^a}{d\s} \frac{dt}{d\s} + \alpha x^a \left(\frac{dt}{d\s}\right)^2 - \frac{\frac{d^2t}{d\s^2}}{\frac{dt}{d\s}}\frac{dx^a}{d\s} = 0\label{invautoeq}
\end{equation}
As we mentioned at the beginning of this section the Nomizu parameters
$\kappa$ and $\iota$ drop out of this equation, so that the geodesic
equation is the same for all invariant connections. Upon choosing time
$t$ to be the parameter along the curve\footnote{Note that if we
  choose $\sigma=t$ then $f=\Gamma^t_{tt}=\kappa+\iota$. So $t$ is not
  an affine parameter if one works with a connection for which
  $\kappa+\iota\neq 0$. Since this sum vanishes for the invariant
  NC-compatible connection \eqref{invcomp}, $t$ is an affine parameter
  in that case.}, i.e. $\sigma=t$, the equation~\eqref{invautoeq}
further simplifies to
\begin{equation}
  \ddot x^a+\beta \dot x^a+\a x^a=0\label{dampeq}
\end{equation}
with $\frac{d}{dt}$ denoted by a dot. Here, we recognize the dynamical
equation of damped harmonic motion, with damping $\beta$ and stiffness
$\alpha$.

We have thus arrived at a somewhat surprising but very concrete
physical interpretation of particle motion on the most general
galilean homogeneous kinematical spacetimes. Just as the invariant
geodesics on Galilei spacetime correspond to free mechanical motion,
the invariant geodesics on the spacetime $\calm_{(\a,\b)}$ correspond
to damped harmonic motion. The invariant geodesics of the torsional
galilean AdS spacetimes correspond to the underdamped cases
($\alpha>\frac{\beta^2}{4}$), while those of the torsional galilean dS
spacetimes are equivalent to the overdamped ones\footnote{Note that in
  these we include the $\alpha<0$ range, where the potential is
  repulsive.} ($\a<\frac{\beta^2}{4}$). The critically damped
oscillator describes the geodesics on the spacetime $\calm_{(1,2)}$,
which is the boundary case $\gamma=1$ of the torsional galilean dS
spacetimes.

\subsection{The kinematical algebra through conserved charges}
\label{sec:charges}

Since the equation \eqref{dampeq} is equivalent to the geodesic
equation of a $\calk_{(\a,\b)}$ invariant connection on
$\calm_{(\a,\b)}$ it will exhibit $\calk_{(\a,\b)}$ symmetry by
construction. To study this symmetry and the corresponding conserved
charges it will be convenient to introduce an action that reproduces
the damped harmonic dynamics \eqref{dampeq} as its Euler-Lagrange
equations. This action, known as the Bateman--Caldirola--Kanai (BCK)
action \cite{bateman1931dissipative, caldirola1941forze, kanai}, is
\begin{equation}
S=\int \frac{m e^{\beta t}}{2}\left( \dot x^a\dot x^a-\alpha x^a x^a\right)dt\label{BCK}
\end{equation}
Here we introduced $m$, the mass of the particle, to provide a proper physical interpretation. 

Since we chose time $t$ as our parameter\footnote{A fully
  reparametrization and coordinate invariant formulation is discussed
  in section \ref{sec:NCbarg}. Although we do not discuss the
  Hamiltonian formulation there, a discussion identical to the one in
  this section can be repeated keeping reparametrization invariance
  manifest, by using the formalism of constrained Hamiltonian
  systems.}, it will be useful to describe transformations in their
passive form. I.e. given a vector field $\xi^\mu$, so that
$\delta x^\mu=\xi^\mu$, we define the passive transformations of the
$x^a(t)$ as
\begin{equation}
\delta_\mathrm{passive}x^a=\xi^a-\dot x^a \xi^t\qquad \delta_{\mathrm{passive}}\dot x^a=\frac{d}{dt}\delta_\mathrm{passive}x^a\qquad \delta_\mathrm{passive}t=0
\end{equation}

The vector fields \eqref{kilvec} that generated the $\mathcal{K}_{(\a,\b)}$ action then lead to the following passive transformations
\begin{equation}
	\delta_{J_{ab}}x^c=x^b\delta_a^c-x^a\delta_b^c\qquad \delta_{P_a}x^b=\dot F(t)\delta_a^b\qquad \delta_{B_a}x^b=F(t)\delta_a^b\qquad \delta_H x^a=-\dot x^a\label{deltaspat}
\end{equation}
with
\begin{equation}\label{eq:defF}
  F(t)=e^{-\beta t/2} \frac{\sin\omega t}{\omega}
\end{equation}
For calculational purposes, it is useful to note that $F$ itself is a
solution to the damped harmonic equation \eqref{dampeq}: $\ddot F+\beta \dot F+\alpha F=0$. One checks that the commutators of the transformations
(\ref{deltaspat}) reproduce those of the vector fields
\eqref{kilvec} and thus also represent the Lie algebra $\mathfrak{k}_{(\a,\b)}$.  

The transformations corresponding to the rotations, boosts and spatial translations are symmetries in the most standard sense, in that they leave the action \eqref{BCK} invariant. Noether's theorem then leads to associated conserved charges. The time translation is a bit more subtle, as it rescales the action with a constant,  i.e. $\delta_H S=\beta S$, rather than leaving it strictly invariant. This is sufficient however to leave the equations of motion invariant and so time translations are indeed, as was warranted by construction, also a symmetry of the particle motion. The association of a conserved charge is now a bit more subtle, but can still be performed. 

The discussion of the conserved charges, as well as the Lie algebra
they form, is clearest in the Hamiltonian formalism. It allows us to
realize the kinematical algebra $\mathfrak{k}_{(\a,\b)}$ as a Lie
algebra of phase space vector fields that can then be made homomorphic
to an algebra of conserved charges. The fact that the time
translations rescale the action rather than leave it invariant (when
$\beta\neq 0$), has two interesting related effects: the algebra of
conserved charges is realized through a slight generalization of the
standard Poisson bracket, and simultaneously the Bargmann extension --
which appears for all Lie algebras $\mathfrak{k}_{(a,b)}$ -- is no
longer central.

First one observes that the canonical momenta defined by the Lagrangian
\eqref{BCK} are
\begin{equation}
	\pc_a=me^{\beta t}\dot x^a\label{momdef}
\end{equation}
and one finds the canonical Hamiltonian
\begin{equation}
	\hc=\frac{e^{-\beta t}}{2m}\pc_a \pc_a+\frac{m\,\alpha\, e^{\beta t}}{2} x^a x^a\label{ham}
\end{equation}
Via \eqref{momdef} one finds the (on-shell) transformations of the
canonical momenta and by introducing the phase space coordinates
$(y^A)=(x^a, \pc_a)$ one can then express the transformations on phase
space in terms of phase space vector fields,
$\Xi=\Xi^a\partial_a+\Xi_a\partial^a$, via $\delta y^A=\Xi^A$. Here we used
the notation $\partial^a=\frac{\partial}{\partial \pc_a}$. Carrying
out this procedure for all symmetry transformations one finds
\begin{eqnarray}
	\Xi_{J_{ab}}&=&x^b\partial_a-x^a\partial_b+\pc_b\partial^a-\pc_a\partial^b\label{XJ}\\
	\Xi_{P_a}&=&\dot F\partial_a+me^{\beta t} \ddot F \partial^a\\
	\Xi_{B_a}&=& F\partial_a+me^{\beta t} \dot F \partial^a\label{XB}\\
	\Xi_{H}&=&-\frac{e^{-\beta t}}{m}\pc_a\partial_a+(m\alpha e^{\beta t}x^a+\beta \pc_a)\partial^a,\label{XH}
\end{eqnarray}
where $F$ is given by equation~\eqref{eq:defF}. One verifies that
under commutation the above phase space vector fields close into an
algebra which is (anti-)isomorphic to the kinematical algebra
$\mathfrak{k}_{(\a,\b)}$ defined in \eqref{isotropybrackets} and
\eqref{galbrackets}.

We should now recall the notion of hamiltonian vector field. The
canonical symplectic form is $\Omega=dx^a\wedge d\pc_a$. A vector field
$X$ is hamiltonian if there exists a phase space function $f$ such
that
\begin{equation}
  i_X \Omega=d f\qquad\Leftrightarrow \qquad X^a=\partial^a f\,,\quad X_a=-\partial_a f
\end{equation}
In this case, we will write $X=X_f$.  It follows that the commutator
of two hamiltonian vector fields can be re-expressed in terms of the
Poisson bracket:
\begin{equation}
  [X_{f_1},X_{f_2}]=-X_{\{f_1,f_2\}}\label{homomorph1}
\end{equation}

In particular, the time evolution is determined by the hamiltonian
vector field associated to the canonical Hamiltonian \eqref{ham}:
\begin{equation}
  X_{\hc}=\frac{e^{-\beta t}}{m}\pc_a\partial_a-m\a e^{\beta t}x^a\partial^a
\end{equation}
via Hamilton's equations
\begin{equation}
  \frac{dy^A}{dt}=X_{\hc}^A
\end{equation}
These imply the following time evolution for an arbitrary
(time-dependent) phase space function $f$:
\begin{equation}
  \frac{d}{dt}f=\partial_t f+\{f,\hc \}\label{symcond1}
\end{equation}
It follows that a phase space function $f$ is conserved when
$\{\hc,f\}=\partial_t f$ or
\begin{equation}
	[X_{f},X_{\hc}]=X_{\partial_t f}\label{conserved}
\end{equation}

The vector fields (\ref{XJ}-\ref{XB}) associated to rotations, spatial
translations and boosts  are all hamiltonian:
\begin{eqnarray}
  \Xi_{J_{ab}}&=&X_{\cJ_{ab}}\qquad\mbox{with}\quad \cJ_{ab}=x^{b}\pc_{a}-x^{a}\pc_{b}\label{ch1}\\
  \Xi_{P_{a}}&=&X_{\cP_{a}}\qquad\mbox{with}\quad \cP_{a}=\dot F \pc_a-me^{\beta t}\ddot F x^a\\
  \Xi_{B_{a}}&=&X_{\cB_{a}}\qquad\mbox{with}\quad \cB_{a}=F \pc_a-me^{\beta t}\dot F x^a,\label{ch2}
\end{eqnarray}
where $F$ is given by equation~\eqref{eq:defF}.  One checks that all  of $\cJ_{ab}$, $\cP_a$ and $\cB_a$ satisfy
\eqref{conserved} and are hence conserved charges. They form the
following Poisson algebra
\begin{eqnarray}
	\{\cJ_{ab},\cJ_{cd}\}&=& \delta_{bc} \cJ_{ad}-\delta_{ac}\cJ_{bd}-\delta_{bd}\cJ_{ac}+\delta_{ad}\cJ_{bc}\label{P1}\\
	\{\cJ_{ab},\cB_c\}&=&\delta_{bc}\cB_a-\delta_{ac}\cB_b\qquad \{\cJ_{ab},\cP_c\}=\delta_{bc}\cP_a-\delta_{ac}\cP_b\label{P2}\\
	\qquad\{\cP_a,\cB_{b}\}&=&m\,\delta_{ab}\label{BM}
\end{eqnarray}
The first two lines, (\ref{P1}, \ref{P2}), reproduce the brackets of
the subalgebra of the kinematical algebra $\mathfrak{k}_{(\a,\b)}$
spanned by rotations, boosts and spatial translations, while the last
line provides the well known Bargmann extension
\cite{bargmann1954unitary}. The mass $m$ is a constant, which implies
that $X_{m}=0$ and that $m$ Poisson commutes with all the $\cJ_{ab}$,
$\cP_a$ and $\cB_a$ -- so that the extension \eqref{BM} is central in
the subalgebra (\ref{P1}, \ref{P2}). It is interesting to note that
although the charges intricately depend on the parameters $(\a,\b)$
and time $t$, see (\ref{ch1}-\ref{ch2}, \ref{eq:defF}), the Bargmann
extension is universal to all cases and independent of these
parameters. In deriving \eqref{BM} the relation
$\dot F^2-\ddot F F=e^{-\beta t}$ is crucial, it expresses the
Wronskian of $F$ and $\dot F$, both of which are solutions to the
dynamical equation \eqref{dampeq}.

We have reserved the discussion of the time translation symmetry and
the associated vector field $\Xi_{H}$ until now, as it is more subtle.
Indeed, the vector field $\Xi_H$ \eqref{XH} is not hamiltonian, indeed
not even symplectic. Instead, it rescales the symplectic structure by
a constant:
\begin{equation}
  \call_{\Xi_H}\Omega=\beta\, \Omega\,.
\end{equation}
We discuss such symplectic homotheties and how they can lead to conserved charges in more generality and detail in
Appendix~\ref{sec:homap}. Here we simply remark that the difference
between $\Xi_H$ and $\beta E$ is however hamiltonian, where we
introduced a preferred symplectic homothety
\begin{equation}
E=\frac{1}{2}(x^a\partial_a+\pc_a\partial^a)\,.\label{explE}
\end{equation} In other words, we can write 
\begin{equation}
	\Xi_H = X_{(\beta,\cH)} := \beta E + X_{\cH}\label{explH}
\end{equation}
where a short computation reveals that
\begin{equation}
	\cH = -\left(\frac{e^{-\beta t}}{2m}\pc_a\pc_a+\frac{\beta}{2}x^a\pc_a+\frac{m \alpha e^{\beta t}}{2} x^a x^a \right) = -\left(\hc+\frac{\beta}{2} x^a \pc_a \right)
\end{equation}
Interestingly $\cH$ is again a conserved charge. This can simply be
verified by direct computation of $\frac{d}{dt}\cH$ and use of the
equations of motion. This is however not a coincidence: in
Appendix~\ref{sec:homap} we show how for any homothetic hamiltonian
vector field $X_{(s,f)}=s E+X_f$ that satisfies
\begin{equation}
\call_{X_{(s,f)}}\hc= s\hc+\partial_t f\qquad \call_E\hc=\hc \qquad \call_E\Omega=\Omega\label{symconds}
\end{equation} 
the phase space function $f$ is a conserved charge. This is a
generalization of the standard argument for hamiltonian vector fields
$X_f=X_{(0,f)}$ to homothetic hamiltonian vector fields $X_{(s,f)}$.
Verifying that indeed \eqref{explE} and \eqref{explH} satisfy the
conditions \eqref{symconds}, with $s=\beta$ and $f=\cH$, is thus another way of showing that $\cH$ is
a conserved charge.

It is now important to point out that although the homothetic
hamiltonian vector fields still form a subalgebra of the algebra of
vector fields (see Appendix~\ref{sec:homap}), this algebra is no longer
homomorphic to the Poisson algebra of phase space functions. Rather
one finds the relation
\begin{equation}
	[X_{(s_1,f_1)},X_{(s_2,f_2)}]=-\dpb{(s_1,f_1)}{(s_2,f_2)}\label{homomorph2}
\end{equation}
where
\begin{equation}
	\dpb{(s_1,f_1)}{(s_2,f_2)}:=\left(0,\{f_1, f_2\}-s_1(\call_E f_2-f_2)+s_2 (\call_E f_1-f_1)\right)\label{dpb}
\end{equation}
It should be clear that \eqref{homomorph2} is a direct generalization
of \eqref{homomorph1} and that furthermore
$\dpb{(0,f_1)}{(0,f_2)}=(0,\{f_1,f_2\})$. Note however that although
$\dpb{\cdot}{\cdot}$ is a Lie bracket, it is \emph{not} a Poisson
bracket (nor even a Jacobi bracket \cite{Kirillov_1976,MR524629}). An
important subtlety that will play a role below, is that the constant
phase space functions $c$, are no longer central (as they are in the
Poisson algebra):
\begin{equation}
  \dpb{(s,f)}{(0,c)}=(0,s c)\,.
\end{equation}

Since it is the algebra of phase space vector fields
(\ref{XJ}-\ref{XH}) that is (anti-) isomorphic to the kinematical
algebra $\mathfrak{k}_{(\a,\b)}$ and since one of these vector fields
is homothetic hamiltonian rather than simply hamiltonian, it follows
that we will be able to recover the kinematical algebra in terms of
conserved charges only with respect to the generalized bracket
\eqref{dpb} and by taking into account the scaling weight $s$, defined
via $\call_\Xi\Omega=s\Omega$, of each symmetry.  Indeed, an explicit
calculation shows that the non-vanishing brackets are
\begin{equation}\label{dP}
  \begin{split}
    \dpb{\mathbf{J}_{ab}}{\mathbf{J}_{cd}} &= \delta_{bc}
    \mathbf{J}_{ad} - \delta_{ac} \mathbf{J}_{bd} - \delta_{bd} \mathbf{J}_{ac} +\delta_{ad} \mathbf{J}_{bc}\\
    \dpb{\mathbf{J}_{ab}}{\mathbf{B}_c} &=
    \delta_{bc}\mathbf{B}_a-\delta_{ac}\mathbf{B}_b\\
    \dpb{\mathbf{J}_{ab}}{\mathbf{P}_c} &= \delta_{bc} \mathbf{P}_a -
    \delta_{ac} \mathbf{ P}_b\\
    \dpb{\mathbf{H}}{\mathbf{B}_a}&= - \mathbf{P}_a\\
    \dpb{\mathbf{H}}{\mathbf{P}_a}&= \alpha \mathbf{B}_a + \beta \mathbf{P}_a\\
    \dpb{\mathbf{P}_a}{\mathbf{B}_{b}} &= \delta_{ab} \mathbf{M}\\
    \dpb{\mathbf{H}}{\mathbf{M}}&=\beta\, \mathbf{M},
      \end{split}
\end{equation}
where we introduced the shorthands
\begin{equation}
  \mathbf{J}_{ab}=(0,\cJ_{ab}), \quad \mathbf{P}_a=(0,\cP_a), \quad
  \mathbf{B}_a=(0,\cB_a), \quad \mathbf{H}=(\beta, \cH)\quad\text{and}\quad \mathbf{M}=(0,m).
\end{equation}
The Lie algebra~\eqref{dP} is a one-dimensional extension, by the
generator $\mathbf{M}$, of the galilean kinematical algebra
$\mathfrak{k}_{(\a,\b)}$ defined in \eqref{isotropybrackets} and
\eqref{galbrackets}. In case $\beta=0$ one recovers the Bargmann
central extensions of the Galilei ($\alpha =0$) and Newton--Hooke
($\alpha>0$ and $\alpha<0$) algebras.  If $\beta\neq 0$, the
appearance of the mass $\mathbf{M}$ on the right hand side of the
bracket of boosts and commutators remains intact, but this Bargmann
extension is now no longer central, as can be seen from the nonzero
$\dpb{\mathbf{H}}{\mathbf{M}}$ bracket. Physically this has the
interpretation of time translations rescaling the mass of the particle
by a constant. Indeed, the physical mass of the particle can be
identified with the overall prefactor of the action. Since
time-translations rescale the action rather than leaving it invariant
when $\beta\neq 0$, the effect of a time translation is indeed to
rescale the physical mass.

For any values of $\alpha,\beta$ the Lie algebras~\eqref{dP} are a
deformation of the centrally extended static kinematical Lie algebra.
Such deformations were classified in
\cite[Table~2]{Figueroa-OFarrill:2017ycu} for $d=3$ and
\cite[Table~18]{Figueroa-OFarrill:2017tcy} for $d>3$. In section
\ref{sec:Eisenhart} we show how they also appear as the algebras of
homotheties of the Lorentzian metrics \eqref{bargstr}, that are
obtained through the Eisenhart lift of the damped harmonic
oscillator. They are also the isometry Lie algebras of certain
homogeneous pp-waves discussed in \cite{Figueroa-OFarrill:2022ryd}, as we explain in appendix \ref{confapp}.

We end this section with a small curiosity. As we discussed above, the conserved charges form a Lie algebra homomorphic to the kinematical algebra only upon the introduction of a modified bracket. Somewhat surprisingly the charges also close under the usual Poisson bracket. In this case they do not reproduce the kinematical algebra $\mathfrak{k}_{(\a,\b)}$, but rather a central extension of the kinematical algebra at {\it different} values of the parameters, i.e. $\mathfrak{k}_{(\omega^2,0)}$. 

To see this explicitly one can first compute
\begin{eqnarray}
 	\{\cH,\cB_a\}&=&-\cP_a-\frac{\beta}{2}\cB_a\qquad \{\cH,\cP_a\}=\alpha \cB_a+\frac{\beta}{2} \cP_a \label{poissonh}
\end{eqnarray}
Upon performing the change of basis
\begin{align}
	\tilde{\cP}_a = \cP_a + \frac{\beta}{2} \cB_a 
\end{align}
one then gets the following Poisson bracket algebra:
\begin{equation}
  \begin{split}
    \{\cJ_{ab},\cJ_{cd}\}&= \delta_{bc} \cJ_{ad}-\delta_{ac}\cJ_{bd}-\delta_{bd}\cJ_{ac}+\delta_{ad}\cJ_{bc}\\
    \{\cJ_{ab},\cB_c\}&= \delta_{bc}b_a-\delta_{ac}b_b\\
    \{\cJ_{ab},\tilde{\cP}_c\} &= \delta_{bc}\tilde{\cP}_a-\delta_{ac}\tilde{\cP}_b\\
    \{\cH,\cB_a\}&= -\tilde{\cP}_a\\
    \{\cH,\tilde{\cP}_a\}&=\omega^2 \cB_a\\
    \{\tilde{\cP}_a,\cB_{b}\}&=m\,\delta_{ab},
  \end{split}
\end{equation}
which is the Bargmann central extension of the galilean kinematical
algebra $\mathfrak{k}_{(\omega^2,0)}$. This shows that the free motion
of a particle on $\calm_{(\a,\b)}$, or equivalently a damped harmonic
oscillator of parameters $\a,\b$, has not only
$\mathfrak{k}_{(\a,\b)}$ symmetry, but also
$\mathfrak{k}_{(\omega^2,0)}$ symmetry. It is well known, see e.g.
\cite{cervero1984quantum}, that the algebra of all symmetries of the
damped harmonic oscillator is larger than $\mathfrak{k}_{(\a,\b)}$,
but what is somewhat surprising is that this larger symmetry group
apparently also has $\mathfrak{k}_{(\omega^2,0)}$ as a subgroup.

\section{Particle motion via Newton-Cartan and Bargmann geometry}
\label{sec:NCbarg}

In the previous section we saw how a free particle, or from a
mathematical point of view a geodesic, on a galilean kinematical
spacetime $\calm_{(\a,\b)}$ is equivalent to damped harmonic
motion. This is made explicit by working in adapted coordinates
$(t,x^a)$ that split time and space and furthermore by choosing time
$t$ as the parameter along the geodesic. In this section we return to
a more intrinsic and geometric description by rewriting the action
\eqref{BCK} in a coordinate and reparametrization invariant way. This
will make use of a Newton-Cartan structure, which -- somewhat
surprisingly -- is not invariant, but only homothetically invariant
under $\calk_{(\a,\b)}$. Indeed, since the action for a particle on a
Newton-Cartan background leads to a compatible symmetric connection,
but the analysis of Section~\ref{sec:invcon} revealed that when
$\beta\neq 0$ the invariant compatible connection is not symmetric, an
action based on the invariant Newton-Cartan structure on
$\calm_{(\a,\b)}$ cannot lead to a geodesic equation involving an
invariant connection (when $\beta\neq 0$).

We will first, in section \ref{sec:particle} review the general
formulation and properties of the action describing particle dynamics
on a Newton-Cartan background and then, in subsection \ref{sec:covab},
specialize to the case of our interest: $\calk_{(\a,\b)}$ invariant
motion on $\calm_{(\a,\b)}$. Finally, we will discuss in subsection
\ref{sec:Eisenhart} how the Bargmann extended algebra, that appeared
through a careful consideration of the conserved charges on phase
space in section \ref{sec:charges}, can be given a geometric
interpretation as an algebra of homotheties\footnote{Actually the same transformations are also isometries of a conformally related metric, see appendix \ref{confapp}.} of an associated (higher
dimensional) Lorentzian metric through the Eisenhart lift.

\subsection{Particle motion on Newton--Cartan spacetimes}
\label{sec:particle}

A Newton--Cartan spacetime is a $d+1$ dimensional manifold $\calm$
equipped with a Newton--Cartan structure $(\tau_\m, h^{\m\n})$, see appendix \ref{sec:type1}. Particle motion on such a Newton--Cartan spacetime is then described in terms of a curve
$x^\mu(\sigma)$, where $x^\m$ are coordinates on $\calm$ and $\sigma$
is a worldline parameter. Free particle motion is essentially geodesic
and so to define it one needs an affine connection. Contrary to the
riemannian or lorentzian case there is more than one torsion free
compatible connection for a given Newton--Cartan structure. This
implies that to define particle motion one needs to provide additional
information. A priory this could be (part of) the connection itself,
but this turns out not to be the most useful way to package this
additional freedom, especially in case one would want to formulate a
variational principle. The action for a particle on a Newton--Cartan
background goes back to \cite{RevModPhys.36.938}, reviews from a more
modern perspective can be found in e.g. \cite{Andringa:2016sot,
	20.500.11850/488630}.

To write a reparametrization invariant particle Lagrangian, one
introduces what we will refer to as a \emph{compatible NC-doublet},
$(\hat \tau^\m, \bar{h}_{\m\n})$, see Appendix~\ref{sec:type1} for a
precise definition and further details.

A Newton--Cartan structure together with a compatible NC-doublet then
define the action
\begin{equation}
	S[x^\m(\sigma)]= \frac{m}{2} \int \frac{\bar{h}_{\m\n}\frac{d x^\m}{d\sigma}\frac{d x^\n}{d\sigma}}{\tau_\r \frac{d x^\r}{d\sigma} }  d\sigma\label{NCaction}
\end{equation}

In this paper we will furthermore restrict attention to the case
$\partial_{[\m}\tau_{\n]}=0$ , the Euler--Lagrange equations for the
action \eqref{NCaction} then take the form
\begin{equation}
	\frac{d^2 x^\m}{d\sigma^2}+\Gamma_{\r\s}^\m \frac{dx^\r}{d\s}\frac{dx^\s}{d\s}=\frac{\frac{d}{d\s}N}{N}\frac{dx^\m}{d\s}\label{NCgeo}
\end{equation}
where $N=\tau_\r \frac{d}{d\s}x^\r$ and
\begin{equation}
	\Gamma^{\lambda}_{\mu\nu}=\hat\tau^\lambda\partial_{(\mu}\tau_{\nu)}+\frac{1}{2}h^{\l\r}(\partial_{\m}\bar h_{\n\rho}+\partial_{\n}\bar h_{\m\rho}-\partial_{\r}\bar h_{\m\n})\label{nccon}
\end{equation}
As the notation suggests, the $\Gamma_{\m\n}^\l$ provide an affine
connection on $\calm$, which is symmetric and leaves the Newton--Cartan
structure $(\tau_\m, h^{\m\n})$ invariant.

We can thus conclude that particle motion on a  Newton--Cartan
spacetime $(\calm, \tau_\m, h^{\m\n})$, as specified by the action
\eqref{NCaction}, is equivalent to geodesic motion with respect to a
symmetric connection compatible with the Newton--Cartan structure. Both
are determined by a choice of  compatible NC-doublet $(\hat \tau^\m,
\bar h_{\m\n})$.

Let us now connect the somewhat abstract discussion above to the more
familiar mechanics of a nonrelativistic particle. Under the
assumption $\partial_{[\m}\tau_{\n]}=0$, we can always (locally) make
a choice of coordinates $(t,x^a)$ such that $\tau_\m=\delta_\m^t$.
Additionally, we can choose the worldline parameter to coincide with
our choice of time: $\sigma=t$.  It follows that
$h^{\m\n}=\delta^\m_a\delta^\n_b h^{ab}$ and we can choose (without
loss of generality) a compatible NC-doublet $\hat
\tau^\m=\delta^\m_t$, $\bar{h}_{\mu\nu}=\delta_\m^a\delta^b_\n
h_{ab}+\delta_\m^t C_\n+\delta_\n^t C_\m$. The action \eqref{NCaction}
then takes the form
\begin{equation}
	S[x^a(t)] = m\int \left(\frac{1}{2}h_{ab}\dot x^a\dot x^b+C_a \dot x^a+C_t \right)dt\label{mechac}
\end{equation}
We recognize here the Lagrangian of a particle moving in a
$d$-dimensional space with Riemannian metric ${h}_{ab}$, under the
influence of a vector potential $C_a$ and scalar potential $-C_0$.

The motion of particles on the galilean
homogeneous spacetimes $\calm_{(\a,\b)}$ are highly symmetric special cases of the above. In general these symmetries, which in this setting  are symmetries of the action
\eqref{NCaction}, are generated by infinitesimal diffeomorphisms on $\calm$ that
take the form
\begin{equation}
	\delta_\xi x^\m=\xi^\m\label{coordtr}
\end{equation}
The Lagrangian \eqref{NCaction} transforms under such a diffeomorphism
as
\begin{align}
	\delta_{\xi} L &= \frac{m}{2}\Bbpl  \frac{(\mathcal{L}_{\xi}\bar h_{\mu\nu})\frac{d x^\mu}{d\s} \frac{dx^\nu}{d\s}}{\tau_\mu \frac{dx^\mu}{d\s}}- \frac{\bar h_{\mu\nu} \frac{dx^\mu}{d\s} \frac{dx^\nu}{d\s} (\mathcal{L}_{\xi}\tau_\rho) \frac{dx^\rho}{d\s}}{(\tau_\mu \frac{dx^\mu}{d\s})^2} \Bbpr \label{variationncl}
\end{align}
This implies that if $\xi^\m$ satisfies
\begin{align}
	\mathcal{L}_{\xi}\tau_\mu &= \zeta \tau_\mu \label{NCtransformation}\\
	\mathcal{L}_{\xi}\bar h_{\mu\nu} &= (\lambda + \zeta) \bar h_{\mu\nu} + 2 \tau_{(\nu}\partial_{\mu)}K\nonumber
\end{align}
for some arbitrary functions $K, \zeta$ and constant $\lambda$,
then
\begin{align}
	\delta_\xi L =  \lambda L + m \frac{d}{d\sigma} K \label{6137}
\end{align}
The above is equivalent to the action \eqref{NCaction} being invariant
up to a rescaling with the constant $\lambda$. It follows that
transformations of the form \eqref{coordtr} with $\xi$ satisfying
\eqref{NCtransformation} leave the equations of motion \eqref{NCgeo}
invariant so can be considered symmetries. The associated conserved
Noether charges are
\begin{align}
	Q_\xi(\sigma) = \frac{\delta L}{\delta \frac{dx^\mu}{d\s}} \xi^\mu -  m K - \lambda \int^\s L d\sigma'
\end{align}
The generalization to symmetries that rescale the action rather than
leave it invariant are crucial, since, as we discussed in  Section~\ref{sec:charges}, the time translations are of this type when $\beta\neq 0$. 

\subsection{Covariant description of motion on $\calm_{(\a,\b)}$}
\label{sec:covab}

We can now view the action \eqref{BCK} as a special case of \eqref{mechac} and that way rewrite it as a manifestly covariant and reparametrization invariant particle action \eqref{NCaction} on a Newton-Cartan spacetime. One verifies that the dynamic equation~\eqref{invautoeq} describing the particle on $\calm_{(\a,\b)}$ equals
 the Euler--Lagrange equations~\eqref{NCgeo} for the action~\eqref{NCaction}
 with Newton--Cartan structure
 \begin{equation}
 \tau_\mu=\delta^t_\mu\qquad h^{\mu\nu} = e^{-\beta t}\delta_a^\mu \delta_a^\nu\label{betaNC}
 \end{equation}
 and compatible NC-doublet
 \begin{equation}
 \hat\tau^\mu =\delta_t^\mu\qquad \bar h_{\mu\nu} = e^{\beta t}\delta^a_\mu \delta^a_\nu-\a e^{\beta t} x^a x^a\delta^t_\mu\delta^t_\nu\label{betatriple}
 \end{equation}
 Observe that the Newton--Cartan structure \eqref{betaNC} is \emph{not}
 the invariant one \eqref{NCstructure} when $\beta\neq0$. The above
 shows however that the \emph{symmetric part}\footnote{Remark that it
 	is only the symmetric part of a connection that appears in the
 	geodesic equation \eqref{autoeq}.} of the unique invariant
      connection \eqref{invcomp} compatible with the
 invariant Newton--Cartan structure \eqref{NCstructure} coincides with
 the symmetric connection \eqref{nccon} compatible with the
 Newton--Cartan structure \eqref{betaNC} and specified by the NC-doublet
 \eqref{betatriple}.
 
 The Newton--Cartan structure \eqref{betaNC} is invariant under the
 subalgebra generated by the $\xi_{P_a}, \xi_{B_a}$ and $\xi_{J_{ab}}$
 of \eqref{kilvec}, but this is not the case for the
 time translations: although $\call_{\xi_H}\tau_\m=0$ one finds that
 $\call_{\xi_H}h^{\m\n}=-\beta h^{\m\n}$. Similarly one verifies that
 the $\xi_{P_a}, \xi_{B_a}$ and $\xi_{J_{ab}}$ satisfy
 \eqref{NCtransformation} for (\ref{betaNC}, \ref{betatriple}) with
 $\lambda = \zeta= 0$ and 
 \begin{equation}
 K_{J_{ab}} = 0\,,\qquad K_{P_a} = e^{\beta t}
 \ddot F x^a\,,\qquad K_{B_a} = e^{\beta t}\dot F x^a
 \,,\label{Ks}
 \end{equation}
 where $F$ was defined in \eqref{eq:defF}. It follows that the transformations
 associated to the rotations, boosts and spatial translations leave the action invariant and are
 thus symmetries in the strictest sense. The time translation
 vector field $\xi_H$ also satisfies \eqref{NCtransformation} for
 (\ref{betaNC}, \ref{betatriple}), but now with
 \begin{equation}
 \lambda_H=\beta\quad\mbox{ and}\quad K_{H}=\zeta_H=0\,.\label{lamH}
 \end{equation} This implies the time-translations do not leave the action
 invariant, but rather rescale it with the constant $\beta$, they are
 therefore symmetries in a slightly weaker sense: they still leave the
 geodesic equation invariant -- as indeed they should since that
 equation was constructed using the invariant connection. I.e. we come, as expected, to the same conclusion as in Section \ref{sec:dho}.

\subsection{Eisenhart lift: the extended kinematical algebra through homotheties}
\label{sec:Eisenhart}

About a century ago Eisenhart \cite{Eisenhart1928_10.2307/1968307}
pointed out that a large class of nonrelativistic mechanical systems
can be equivalently described in terms of null geodesic motion on a
higher dimensional lorentzian spacetime with a null Killing vector. This
relation was rediscovered by Duval et al. \cite{Duval:1984cj} (see also \cite{tulczyjew1985intrinsic,omote1989galilean,omote1989galilean2}) and more recently studied from a more general perspective in \cite{Bekaert:2013fta, Morand:2018tke}, among others.  Looked at from a top down perspective this equivalence becomes a null reduction
\cite{Duval:1984cj, Julia:1994bs}.

From a geometric perspective, and restricted to the subclass of cases
that we are interested in, the relation can be precisely phrased as
the equivalence of $d+2$ dimensional Bargmann structures with parallel
null Killing vector and $d+1$ dimensional torsionless Type I
Newton--Cartan geometry.

Let us for clarity recall the relevant definitions. A Bargmann
structure is a lorentzian metric $g_{AB}$ together with a nowhere
vanishing null vector field $k^A$. This vector field is parallel when
$\nabla_A k^B=0$ (w.r.t. the Levi-Civita connection of $g_{AB}$) and
is then also automatically Killing. Type I Newton--Cartan geometry can
take various equivalent shapes. We'll define it to be a Newton--Cartan structure $(\tau_\m, h^{\r\s})$ together with an
equivalence class of compatible NC doublets $[(\hat\tau^\mu,\bar
h_{\r\s})]$, see Appendix~\ref{sec:type1} for more details, it is
torsionless when the intrinsic torsion
\cite{Figueroa-OFarrill:2020gpr} of the galilean $\calg$-structure
vanishes, i.e. $d\tau=0$.

The equivalence between these two geometric concepts is most explicit
upon a choice of coordinates $(x^A)=(u,x^\m)$ such that
$k=\partial_u$. The $(d+2)$ dimensional lorentzian metric then takes
the form
\begin{equation}
ds^2=-2 du \tau_\mu dx^\m+\bar h_{\m\n}dx^\m dx^\n\label{nullred}
\end{equation}
In components this implies
\begin{equation}
(g_{AB})=\begin{pmatrix}
0 &-\tau_\m\\
-\tau_\n & \bar h_{\m\n}
\end{pmatrix}\qquad (g^{AB})=\begin{pmatrix}
-\hat{\tau}^\r\hat{\tau}^\s \bar h_{\r\s} & -\hat \tau^\m\\
-\hat\tau^\n & h^{\m\n}
&
\end{pmatrix} 
\end{equation}
A different choice of coordinates $u'=u+\Lambda(x^\m)$ is equivalent to a change
\begin{equation}
\hat{\tau}^{'\mu}=\hat{\tau}^\m-h^{\m\n}\partial_\m\Lambda\qquad \bar{h}'_{\m\n}=\bar h_{\m\n}+\partial_\m\Lambda\tau_\n+\partial_\n\Lambda\tau_\m\label{doubletredef}
\end{equation}
which leaves the equivalence class invariant, i.e.
$[(\hat{\tau}^\m,\bar{h}_{\m\n})]=[(\hat{\tau}^{'\m},\bar{h}'_{\m\n})]$.
Note that the identification \eqref{nullred} implies that
$(k_A)=(0,\tau_\m)$ and so $k$ being parallel becomes equivalent to
$\tau$ being closed.

Eisenhart's equivalence between nonrelativistic motion and null
geodesic motion is then a direct consequence of the geometric
equivalence reviewed above. First, one observes that a null curve
$(x^A(s)) = (u(s),x^\mu(s))$ for the metric \eqref{nullred} satisfies
\begin{equation}
\dot u=\frac{1}{2}\frac{\bar h_{\m\n}\dot x^\mu \dot x^\nu}{\tau_\rho \dot x^\rho}\label{nullcond}
\end{equation}
Using this relation (and that $\tau$ is closed), a short calculation
then reveals that for such a null curve
\begin{equation}
\ddot x^A+\Gamma^A_{BC}\dot x^A\dot x^C=f\, \dot
x^A\quad\Leftrightarrow\quad \ddot x^\mu+\Gamma^\m_{\r\s}\dot x^\r\dot
x^\s=\frac{\dot N}{N}\, \dot x^\mu 
\end{equation}
On the left side of the equivalence the connection is the Levi-Civita
connection of the metric $g_{AB}$, while on the right hand side the
connection is the Newton--Cartan one \eqref{nccon}, and one recognizes
the equation \eqref{NCgeo} which, as discussed in section
\ref{sec:particle}, via the choices $(x^{\m})=(t,x^i)$, $s=t$ describes
the motion of a nonrelativistic particle described by the action
\eqref{mechac}. Interestingly, the relation \eqref{nullcond} also
provides a geometric interpretation of the NC action \eqref{NCaction},
i.e. $S_{\mathrm{NC}}=\Delta u$, or in other words: Nonrelativistic
particle motion follows null geodesics upon the Eisenhart lift and furthermore such null geodesics extremize the coordinate difference $\Delta u$ between start and endpoint. Note that this is
due to the Bargmann structure exhibited by the lifted geometry and
remark that $\Delta u=\Delta \tilde u$ is indeed a geometric invariant
(although we expressed it in local coordinates).

Possibly the most interesting feature of the Eisenhart lift is
that it provides a new, more geometric, perspective on the symmetries
of the mechanical system \cite{Duval:1984cj}. To start the discussion we
recall that the right-hand side of \eqref{nullcond} is the
nonrelativistic Lagrangian, i.e. $m \dot u=L$. Those symmetries of the
lower dimensional nonrelativistic system that leave the Lagrangian
invariant only up to the addition of a total derivative, i.e. $\delta
L=m \dot K$, will thus have to act on $u$ as well. In particular, to
leave the relation \eqref{nullcond} invariant one has to impose
$\delta u=K$ . If we let $\xi^\mu$ be the $(d+1)$ dimensional
vector field generating the nonrelativistic symmetry, i.e. $\delta
x^\m=\xi^\mu$, then its lift to $(d+2)$ dimensions will thus be
$(\hat\xi^A)=(K,\xi^\mu)$. It follows that the $(d+2)$ dimensional
bracket is
$([\hat\xi_1,\hat\xi_2]_{d+2}^A)=(\xi_1K_2-\xi_2K_1,[\xi_1,\xi_2]^\mu_{d+1})$.
The $(d+1)$ dimensional vector fields form by assumption an algebra,
i.e. $[\xi_1,\xi_2]_{d+1}=\xi_3$, and it thus follows that\footnote{Note that, by definition, $\hat \xi_3 = K_3 \partial_u + \xi_3$. But $a$ is defined in a way that it
	contains $-K_3$ and thus $\hat \xi_3 + a k$ is independent of
	$K_3$.}
\begin{equation}
[\hat\xi_1,\hat \xi_2]_{d+2}=\hat \xi_3+a\, k\qquad a=(\xi_1 K_2-\xi_2 K_1-K_3),\ \xi_3=[\xi_1,\xi_2]_{d+1}\label{liftalg1}
\end{equation} 
So if there are symmetries of the original nonrelativistic mechanical
system that change the Lagrangian by a total derivative, then through
the Eisenhart lift they extend the symmetry algebra by an extra
generator $a k$.\footnote{Let us remark that since $k=\partial_u$ is a
  Killing vector, it generates a symmetry of the (null) geodesic
  equation and is thus automatically a $(d+2)$ dimensional symmetry.
  In case $a$ is not constant then $ak$ is a generator linearly
  independent of $k$ and in that case the full $(d+2)$ dimensional
  algebra is an extension of the $(d+1)$ dimensional algebra by at
  least $a k$ and $k$. In this paper we will focus only on the
  extension by $ak$ only.}

A striking example is already provided by the Galilei symmetry algebra
$\mathfrak{k}_{(0,0)}$ of a nonrelativistic free particle. Taking
$\xi_1=\xi_P=\partial_x$ to be a translation and
$\xi_2=\xi_B=t\partial_x$ a boost, one finds that $\xi_3=0$, as well as
$K_1=K_3=0$ but $K_2=x$. It follows that $[\hat \xi_P ,\hat
\xi_B]_{d+2}= k$. Furthermore $k$ commutes with all other generators
$\hat \xi$ of the (lifted) Galilei symmetries and one thus recovers the
Bargmann extension of the Galilei group. The Eisenhart lift thus
provides a geometric realization of this classic central extension.

We are now ready to return to the main topic of this paper, namely the
galilean kinematical algebras of generic type. In that case, as we
discussed in Section~\ref{sec:charges} and Section~\ref{sec:covab}, we
need to take into account that some symmetries (in particular time
translation) are of an even more general type, in that they also
rescale the Lagrangian rather than only leaving it invariant up to a
total derivative, i.e. $\delta L=\lambda L+m\dot K$. Via
\eqref{nullcond}, this then implies $\delta u=\lambda u+K$ which leads
to a more general form of the lift, i.e.,
\begin{equation}
\hat \xi=(\lambda u+K)\partial_u+\xi\label{hatxi}
\end{equation}
As a check, one verifies that if $\xi$ is a symmetry of the
nonrelativistic dynamics, i.e. satisfies \eqref{symconds}, then $\hat
\xi$ is a conformal Killing vector of the Bargmann structure and hence a symmetry
of the null-geodesic equation, i.e.
\begin{equation}
\call_{\hat \xi}g_{AB}=(\lambda+\zeta) g_{AB}\,.\label{gtransfo}
\end{equation}
Let us recall that, as mentioned in the previous subsection, $\zeta$ is zero for all vector fields \eqref{kilvec} generating $\mathfrak{k}_{(\a,\b)}$. In this case the lifted vector fields $\hat \xi$ then act as homotheties, since $\lambda$ is constant. 

The algebra of the lifted vector fields is then
\begin{equation}
[\hat\xi_1,\hat \xi_2]_{d+2}=\hat \xi_3+a\, k\qquad a=(K_1\lambda_2-K_2\lambda_1+\call_{\xi_1} K_2-\call_{\xi_2} K_1-K_3),\ \xi_3=[\xi_1,\xi_2]_{d+1}\label{liftalg2}
\end{equation}
So apart from the small addition to the definition of $a$ with respect
to \eqref{liftalg1}, it might appear not much has changed. But that
is not the case. Let us restrict attention to the case where $a$ is
constant, which is the case of our interest, so that \eqref{liftalg2}
is an extension of the symmetries by the generator $k$. Now, observe
that via \eqref{hatxi} we have $[k,\hat{\xi}]_{d+2}=\lambda k$. This
implies that $k$ does not commute with those symmetries, $\hat \xi$,
that rescale the nonrelativistic Lagrangian and so when such
symmetries are present, the extension provided by the Eisenhart
lift is no longer central!

The whole discussion above is made explicit in the examples provided
by the kinematical homogeneous spacetimes of galilean type that we
have been studying in this paper. The Type I Newton--Cartan geometry
(\ref{betaNC}, \ref{betatriple}) describing the particle motion
corresponds to the Bargmann structure
\begin{equation}
ds^2=-2du dt-\a e^{\beta t}(x^ax^a)dt^2+e^{\beta t}dx^a dx^a\qquad k=\partial_u\label{bargstr}
\end{equation}
In particular, via the general formalism reviewed above, this implies
that the null-geodesic equations of this lorentzian metric are
equivalent to the equations of motion of the damped harmonic
oscillator \eqref{dampeq}, as indeed can be verified explicitly.
Furthermore, using \eqref{Ks} and \eqref{lamH}, the vector fields
\eqref{kilvec}, that form the kinematical algebra
$\mathfrak{k}_{(\a,\b)}$ and that are symmetries of the
nonrelativistic system, get lifted to
\begin{equation}\label{liftkilvec}
  \begin{split}
    \hat\xi_{J_{ab}} &=\xi_{J_{ab}} \\
    \hat\xi_{H} &=\beta u \partial_u+\xi_H \\
    \hat\xi_{P_{a}} &=e^{\beta t}x^a\ddot{F}\partial_u+\xi_{P_a} \\
  \hat\xi_{B_{a}} &=e^{\beta t}x^a\dot{F}\partial_u+\xi_{B_{a}},
\end{split}
\end{equation}
where $F$ was defined in equation~\eqref{eq:defF}.  One verifies that
these are indeed homothetic Killing vectors of the metric
\eqref{bargstr}. Together with the Killing vector
\begin{equation}
  \hat \xi_M=-\partial_u
\end{equation}
they form the Lie algebra
\begin{equation}
  \begin{split}
    [\hat \xi_{J_{ab}},\hat \xi_{J_{cd}}]&= -\delta_{bc}\hat
    \xi_{J_{ad}}+\delta_{ac}\hat \xi_{J_{bd}}+\delta_{bd}\hat \xi_{J_{ac}}-\delta_{ad}\hat \xi_{J_{bc}}\\
    [\hat \xi_{J_{ab}},\hat \xi_{B_c}]&= -\delta_{bc}\hat
    \xi_{B_a}+\delta_{ac}\hat \xi_{B_b}\\
    [\hat \xi_{J_{ab}},\hat \xi_{P_c}]&=-\delta_{bc}\hat \xi_{P_a}+\delta_{ac}\hat \xi_{P_b}\\
    [\hat \xi_{H},\hat \xi_{B_a}]&=\hat \xi_{P_a}\\
    [\hat \xi_{H},\hat \xi_{P_a}]&=-\alpha\hat \xi_{B_a}- \beta\hat \xi_{P_a}\\
    [\hat \xi_{P_a},\hat \xi_{B_b}]&=-\delta_{ab}\hat \xi_{M}\\
    [\hat \xi_H,\hat{\xi}_M]&=-\beta \hat \xi_M.
  \end{split}\label{lalg}
\end{equation}
This algebra is (anti-)isomorphic to the one formed by the conserved
charges \eqref{dP} and provides the same extension of the kinematical
algebra $\mathfrak{k}_{(\a,\b)}$. In conclusion, we see that the Eisenhart
lift provides a geometric realization, namely as homotheties of the
metric \eqref{bargstr}, of this (non-centrally) extended algebra in a
direct generalization of what happens for the usual free particle. The
universality of this extension suggests to refer to it as the Bargmann
extension of the galilean kinematical algebra
$\mathfrak{k}_{(\a,\b)}$, for any value of $\a$ and $\beta$.

\section*{Acknowledgements}

It is a pleasure to thank Andrew Beckett, Eric Bergshoeff, Mahmut
Elbistan, Joaquim Gomis, Ross Grassie, Stefan Prohazka and Diederik
Roest for useful discussions and conversations. CG is supported by the
2209-A program of TÜBİTAK under grant number 1919B012106282 and DVdB
is partially supported by Boğaziçi University Research Fund under
grant number 21BP2.

\appendix

\section{Type I Newton--Cartan geometry: doublets, triplets and connections}
\label{sec:type1}

Given a $(d+1)$ dimensional manifold $\calm$, a
Newton--Cartan\footnote{Throughout the literature (slightly) different
  nomenclature has been used. As reviewed in
  \cite{Figueroa-OFarrill:2020gpr} the pair $(\tau_\m,h^{\m\n})$ are
  (the local components of characteristic tensor fields of) a
  $\mathrm{Gal}_0$-structure. Here $\mathrm{Gal}_0$ is the homogeneous
  Galilei group that can be defined as the subgroup of
  $\mathrm{GL}(d+1,\mathbb{R})$ consisting of the matrices$  
  \begin{pmatrix}
    1 & 0^T\\ v & A
  \end{pmatrix} \in \GL(d+1,\RR)$, with $v \in
  \RR^d, A \in \Ort(d). $ Some authors for this reason speak of
  a Galilei (or galilean) structure, as in
  e.g. \cite{Duval:1984cj} and reserve the notion
  Newton(-Cartan) structure for a Galilei structure together
  with a (particular type of) compatible connection. In
  \cite{Bergshoeff:2022eog} a similar distinction is made by
  referring to a Galilei structure as a weak Newton-Cartan
  structure. In this paper we will however use the term
  Newton-Cartan structure interchangeably for Galilei, galilean
  or weak Newton-Cartan structure, i.e. to refer to the pair
  $(\tau_\m,h^{\m\n})$, without any additional notions
  implied. Finally let us point out that in
  e.g. \cite{Bekaert:2014bwa} the term leibnizian structure is
  used.} 
structure is a pair of tensors $(\tau_\m,h^{\m\n})$, where $\tau_\mu$
everywhere spans the kernel of $h^{\m\n}$, which is symmetric and
positive semi-definite. See e.g. \cite{Bergshoeff:2022eog} for a
modern review.

For each Newton-Cartan structure one can define the following two additional notions:
\begin{itemize}
\item a \emph{compatible NC-doublet} is an equivalence class $[(\hat
  \tau^\m, \bar h_{\m\n})]$ of tensors $(\hat \tau^\m, \bar h_{\m\n})$ such that
  \begin{equation}\label{eq:nc-doublet}
    \tau_\m\hat \tau^\n+\bar h_{\m\r}h^{\r\n}=\delta_\m^\n
  \end{equation}
  with the equivalence relation $(\hat \tau^\m, \bar
  h_{\m\n})\sim(\hat \tau^{'\m}, \bar h'_{\m\n})$ if and only if
  \begin{eqnarray}\label{eq:eq-relation-nc-doublet}
    \hat\tau^{'\mu}&=&\hat\tau^\m-h^{\m\n}\partial_\n\Lambda \label{deq1}\\
    \bar h_{\m\n}'&=&\bar h_{\m\n}+\tau_\m\partial_\n\Lambda+\tau_\n\partial_\m\Lambda\label{deq2}
  \end{eqnarray}

\item a \emph{compatible NC-triplet} is an equivalence class
  $[(\tau^\m, h_{\m\n}, C_\m)]$ of tensors $(\tau^\m, h_{\m\n}, C_\m)$
  such that
  \begin{equation}
    \tau_\m \tau^\n+h_{\m\r}h^{\r\n}=\delta_\m^\n\qquad \tau^\m\tau^\n h_{\m\n}=0\label{tcomp}
  \end{equation}
  with the equivalence relation $(\tau^\m, h_{\m\n}, C_\m) \sim
  (\tau^{'\m}, h'_{\m\n}, C'_\m)$ if and only if
  \begin{eqnarray}
    \tau_\m \chi^\m&=&0\\
    \tau^{'\mu}&=&\tau^\m-\chi^\m\label{teq1}\\
    h'_{\m\n}&=&h_{\m\n}+h_{\m\r}\chi^\r\tau_\n+h_{\n\r}\chi^\r\tau_\m+h_{\r\s}\chi^\r\chi^\s\tau_\m\tau_\n\label{teq2}\\
    C'_\m&=&C_\m-h_{\m\r}\chi^\r-\frac{1}{2}h_{\r\s}\chi^\r\chi^\s \tau_\m+\partial_\m\Lambda\label{teq3}
  \end{eqnarray}
\end{itemize}

The notions of compatible NC-triplet and compatible NC-doublet are
actually equivalent. This can be verified by considering the following
explicit maps between them
\begin{align}
[(\tau^\mu,h_{\mu\nu},C_\mu)] \mapsto [(\hat{\tau}^\mu, \bar{h}_{\mu\nu})]= [(\tau^\mu-h^{\mu\nu}C_\nu,h_{\mu\nu}+ \tau_\mu C_\nu + \tau_\nu C_\mu)] \label{maptd}\\
[(\hat{\tau}^\mu, \bar{h}_{\mu\nu})] \mapsto[(\tau^\mu,h_{\mu\nu},C_\mu)] = [(\hat{\tau}^\mu,\bar{h}_{\mu\nu} - \tau_\mu\tau_\nu \hat{\tau}^\rho\hat{\tau}^\sigma \bar{h}_{\rho\sigma},-\frac{1}{2}\tau_\mu \hat{\tau}^\rho\hat{\tau}^\sigma \bar{h}_{\rho\sigma})]
\end{align}

A Newton--Cartan structure together with a choice of compatible
NC-triplet or -doublet was dubbed \emph{Type I Newton--Cartan geometry}
in \cite{Hansen:2019pkl}, see also \cite{Hansen:2020pqs,
  20.500.11850/488630}. As we review in Section~\ref{sec:Eisenhart} this
geometric structure is also equivalent to a Bargmann structure in one
dimension higher.

The first use of NC-doublets/triplets is that they define an affine
connection compatible with the Newton--Cartan structure. If we assume
the intrinsic torsion of the NC structure to vanish, i.e. $d\tau=0$,
then this connection takes the form
\begin{eqnarray}
\Gamma_{\m\n}^\rho&=&\hat\tau^\lambda\partial_{\mu}\tau_{\nu}+\frac{1}{2}h^{\l\r}(\partial_{\m}\bar h_{\n\rho}+\partial_{\n}\bar h_{\m\rho}-\partial_{\r}\bar h_{\m\n})\label{con1}\\
&=&\tau^\rho \partial_\mu \tau_\nu + \frac{1}{2} h^{\rho\sigma}(\partial_\mu h_{\nu\sigma}+\partial_\nu h_{\mu\sigma} - \partial_\sigma h_{\mu\nu}) - h^{\rho\sigma}K_{\sigma(\mu}\tau_{\nu)}\label{con2}
\end{eqnarray}
where
\begin{equation}
K_{\m\n}=\partial_\m C_\n-\partial_\n C_\m
\end{equation}
One can directly verify that the connection above is invariant under
the equivalences (\ref{deq1}-\ref{deq2}) and (\ref{teq1}-\ref{teq3})
and that \eqref{con1} and \eqref{con2} are related via \eqref{maptd}. 
Essentially\footnote{A subtle exception is provided by connections of
  the form \ref{con2} where $K_{\m\n}$ is a two form that is not
  closed.} all symmetric compatible connections are of the form
(\ref{con1}, \ref{con2}), see e.g.  \cite{ Hartong:2015zia,
  Bekaert:2014bwa}. 

A second, related, application is that a choice of NC-doublet/triplet
allows to write a Lagrangian for a particle on a manifold with Newton-Cartan structure:
\begin{equation}
S=\int \frac{1}{2}\frac{\bar h_{\m\n}\dot x^\m \dot x^\n}{\tau_\m \dot
  x^\m}d\sigma=\int\left( \frac{1}{2}\frac{h_{\m\n}\dot x^\m \dot
    x^\n}{\tau_\m \dot x^\m}+C_\m \dot x^\m\right) d\sigma
\end{equation}
Again it is straightforward to check that this action does not depend
on a change of representative (\ref{deq1}-\ref{deq2}) and
(\ref{teq1}-\ref{teq3}). The Euler--Lagrange equations of this action
describe geodesic motion with respect to the affine connection
(\ref{con1}, \ref{con2}), see \eqref{NCgeo}.

\section{Symplectic Homotheties}
\label{sec:homap}

Symmetries of the equations of motion are not necessarily symmetries
of the action in the strict sense of leaving it invariant, rather they
can also rescale the action with a constant. Using the term
\emph{homothety} for a diffeomorphism that multiplies a tensor with a
constant we can rephrase the previous statement as saying that
homotheties of the action map solutions of the equation of motion to
other solutions. These symmetries in the more general sense have been
considered in e.g.,  \cite{van1972transformation,
  MR356131, nachtergaele1986groups, carinena2012geometric,
  MR3078679, Zhang:2019koe} (in a gravity context they are
sometimes referred to as trombone symmetries \cite{Cremmer:1997xj}),
and when considered on phase space they are generically
\emph{symplectic homotheties} (i.e. homotheties of the symplectic
form) rather than symplectomorphisms.

Since results on symplectic homotheties appear somewhat scattered in
the literature we collect some of the key concepts and their
properties in this appendix. In addition, we will discuss the
generalization of the extension of the Lie algebra of hamiltonian
vector fields to the Poisson algebra of phase space functions to the
homothetic case, where this extension is no longer central. This last
result has not previously appeared in the literature as far as we are
aware.

In this appendix $(\cals,\Omega)$ will be assumed to be a symplectic
manifold; that is $\Omega$ is a closed non-degenerate two-form on
$\cals$. One then defines a \emph{homothetic symplectic vector field}
as a vector field $X$ such that
\begin{equation}
\call_X\Omega=s\, \Omega\,,\quad s\in\mathbb{R}\,.\label{sympthetydef}
\end{equation}
Given two such vector fields $X,Y$ one verifies
$\call_{[X,Y]}\Omega=0$ and thus the homothetic symplectic vector
fields form a Lie algebra which we will denote as
$\mathfrak{sym}(\cals)$. The definition defines a map $\sigma:\,
X\mapsto s=\sigma(X)$ which is a Lie algebra homomorphism. In
particular the kernel of this map, i.e. those vector fields $X$ for
which $s=0$, are the symplectic vector fields that form the Lie
algebra $\mathfrak{sym}_0(\cals)$. Vector fields with $s\neq 0$ exist
iff  the symplectic form $\Omega$ is exact, i.e. $\Omega=d\theta$,
since then $\Omega=\frac{1}{s}d(i_X\Omega)$. Note that this is the
case for the canonical symplectic form on a cotangent bundle. It
follows from the above that on an exact symplectic manifold $\cals$
one has $\mathfrak{sym}{(\cals)}/\mathfrak{sym}_0(\cals)=\mathbb{R}$.

Before we continue let us recall that for a symplectic vector field $X$
the one-form $i_X\Omega$ is closed, and that when $i_X\Omega$ is
furthermore exact $X$ is called {\it hamiltonian}. Such vector fields form a
subalgebra $\mathfrak{ham}\subset \mathfrak{sym}_0$. The
non-degeneracy of the symplectic form allows to associate to every
real function $f$ on $\cals$ a hamiltonian vector field $X_f$ via
\begin{equation}
i_{X_f}\Omega=df
\end{equation}  
The map $f\mapsto X_f$ is a Lie algebra (anti-)homomorphism if we
equip the space of functions with the Poisson bracket
$\{f,g\}=\Omega(X_f,X_g)$, i.e.,
\begin{equation}
  [X_f,X_g]=X_{-\{f,g\}}\,.
\end{equation}
Since the kernel of the map $f\mapsto X_f$ are the constant functions,
which are central with respect to the Poisson bracket, one observes
the well known fact that the Poisson algebra of functions is a central
extension of the Lie algebra of hamiltonian vector fields.

Let us now get back to the homothetic generalization. Assuming from
now on that $\Omega=d\theta$ is exact we can define an\footnote{Note
  that the choice of the symplectic potential $\theta$, and hence the
  associated Euler vector field, is not unique. The possible choices
  differ by an arbitrary closed form. So although one could choose to
  write $E_\theta$ to emphasize the dependence on the choice of
  $\theta$ we will refrain from doing so to ease notation.} {\it Euler
vector field} $E$ via\footnote{Since $\Omega$ is non-degenerate $E$ is
  unique given $\theta$.}
\begin{equation}
i_{E}\Omega=\theta
\end{equation}
Now remark that $X-\sigma(X) E$ is symplectic when $X$ is homothetic
symplectic. If furthermore $X-\sigma(X) E$ is hamiltonian we say that
$X$ is \emph{homothetic hamiltonian} (with respect to $\theta$). It
follows directly from this definition that for every such vector field
there exists a function $f$ and a real number $s$ such that
$X=X_{(s,f)}$ where
\begin{equation}
X_{(s,f)}=s E+X_f\label{Xcfdef}
\end{equation}
A short computation reveals that
\begin{equation}
[X_{(s_1,f_1)},X_{(s_2,f_2)}]=X_{-\dpb{(s_1,f_1)}{(s_2,f_2)}}
\end{equation}
where
\begin{equation}
\dpb{(s_1,f_2)}{(s_2,f_2)}=\left(0,\{f_1,f_2\}+s_1 (f_2-E[f_2])-s_2(f_1-E[f_1])\right)
\end{equation}
First of all this implies that the homothetic hamiltonian vector
fields form a Lie algebra, that we denote as\footnote{Indeed different
  choices of symplectic potential $\theta$ lead to different algebras
  $\mathfrak{ham}_\theta(\cals)$, these are isomorphic iff the two
  potentials are cohomologous.}
$\mathfrak{ham}_\theta(\cals)$. It also makes clear that
$\mathfrak{ham}(\cals)$ is an ideal in $\mathfrak{ham}_\theta(\cals)$.
Indeed, one has the Lie algebra extension by derivation $0\rightarrow
\mathfrak{ham}(\cals)\rightarrow\mathfrak{ham}_\theta(\cals)\rightarrow
\mathbb{R}\rightarrow 0$ \cite{MR356131}. Furthermore
$\dpb{\cdot}{\cdot}$ defines a Lie\footnote{Note that although
  $\dpb{\cdot}{\cdot}$ is a Lie bracket, it is neither a Poisson nor
  Jacobi bracket.} algebra on the space
$C^{\infty}_\theta(\cals)=\mathbb{R}\oplus C^{\infty}(\cals)$. By
construction the map $(s,f)\mapsto X_{(s,f)}$ is a Lie algebra
homomorphism, and via \eqref{Xcfdef} one infers the kernel is given by
$(0,c)$ with $c\in\mathbb{R}\subset C^\infty(\cals)$ being a constant
function. So also in the homothetic generalization it remains true
that $C_\theta^\infty(\cals)$ is a one-dimensional extension of
$\mathfrak{ham}_\theta(\cals)$, but it is no longer true that this
extension is central, since
\begin{equation}
\dpb{(0,c)}{(s,f)}=(0,-sc)
\end{equation}
A summary of the various Lie algebra extensions is provided below:
\begin{equation*}
  \begin{tikzcd}
    & & 0 \arrow[d] & 0 \arrow[d] & \\
    0 \arrow[r] & \RR \arrow[d,equal] \arrow[r] & C^\infty(M) \arrow[d] \arrow[r] & \mathfrak{ham}(M) \arrow[d] \arrow[r] & 0\\
    0 \arrow[r] & \RR \arrow[r] & C_\theta^\infty(M) \arrow[d] \arrow[r] & \mathfrak{ham}_\theta(M) \arrow[d]
    \arrow[r] & 0\\
    & & \RR \arrow[d] \arrow[r,equal] & \RR \arrow[d] & \\
    & & 0 & 0 & \\
  \end{tikzcd}
\end{equation*}
All arrows in the above commutative diagram are Lie algebra
homomorphisms. The four Lie algebra extensions discussed in the text
correspond to the two vertical and two horizontal short exact
sequences.

In a physical context, the homothetic symplectic vector fields are
only relevant when, in addition, they leave invariant Hamilton's
equations.  This extra condition can alternatively be expressed by
considering the hamiltonian action
\begin{equation}
  S[\gamma]=-\int_{t_\mathrm{i}}^{t_\mathrm{f}} (\gamma^*\theta+\gamma^*\hc\, dt)\,,
\end{equation}
with $\gamma$ a curve and $\hc$, the Hamiltonian, a function on $\cals$, and requiring
that under $\delta\gamma=X_{(s,f)}$
\begin{equation}
  \delta S=\lambda S+q(\gamma_\mathrm{i},\gamma_\mathrm{f})\label{allowedtransfo}
\end{equation}
where $q$ is some arbitrary function of the endpoints of the curve.

Define now (the generator of) a \emph{symmetry} of an exact
hamiltonian system $(\cals,\theta, \hc)$ to be a homothetic hamiltonian
vector field $X_{(s,f)}$ such that\footnote{Note that
  \eqref{symcondgen} can equivalently be rewritten as
  $\call_{X_{(s,f)}} X_{\hc}+X_{\partial_t f}=0$ or
  $\dpb{(0,\hc)}{(s,f)}=(0,\partial_t f)$.}
\begin{eqnarray}
\call_{X_{(s,f)}}\hc=s\, \hc+\partial_t f\label{symcondgen}
\end{eqnarray}
It then follows that under such symmetries indeed
\eqref{allowedtransfo} is guaranteed. To verify this, remark that
\begin{equation}
\delta \int_{t_\mathrm{i}}^{t_\mathrm{f}} \gamma^*\hc\, dt=\int_{t_\mathrm{i}}^{t_\mathrm{f}} \gamma^*(\call_{X_{(s,f)}}\hc)\, dt
\end{equation}
while
\begin{equation}
\delta \int_{t_\mathrm{i}}^{t_\mathrm{f}} \gamma^*\theta=s\int_{t_\mathrm{i}}^{t_\mathrm{f}} \gamma^*\theta+\left.(f-\call_E f)\right|_{\gamma_\mathrm{i}}^{\gamma_{\mathrm{f}}}-\int_{t_\mathrm{i}}^{t_\mathrm{f}} \gamma^*\partial_t f\, dt
\end{equation}
Here we used that\footnote{Remark:
  $i_{X_f}\theta=i_{X_{f}}i_E\Omega=-i_E df=-\call_Ef$ and
  $i_E\theta=i_E i_E\Omega=0$.}
\begin{equation}
\delta (\gamma^* \theta)=\gamma^* \call_{X_{(s,f)}}\theta-\gamma^*(\partial_t\call_E f)\quad\mbox{and}\quad
\call_{X_{(s,f)}}\theta=s\,\theta+d(f-\call_Ef)
\end{equation}
Now observe
\begin{eqnarray}
	\frac{d}{dt}f&=&\call_{X_\hc}f+\partial_t f\\
	&=&-\call_{X_{(s,f)}}\hc+s\call_E\hc+\partial_t f\\
	&=&s(\call_E \hc-\hc)
\end{eqnarray}
We thus see that a symmetry $X_{(s,f)}$ leads to a conserved charge $f$ when
\begin{equation}
s=0\quad\mbox{or}\quad \call_E \hc=\hc\label{Econd}
\end{equation}
We should point out that the condition \eqref{Econd} depends on the
choice of Euler vector field (and thus choice of symplectic potential
$\theta$).

In summary, for the existence of a conserved charge $f$ it is thus
sufficient that there exists an Euler vector field for which both
\eqref{Econd} and \eqref{symcondgen} hold.

\section{Low dimensional cases}
\label{sec:lowdap}

When the number of spatial dimension is two or less, $d\leq 2$, there
appear some exceptions to the discussion of $d>2$ in the main text.
When $d=2$ there exists an additional kinematical algebra of galilean
type \cite{Figueroa-OFarrill:2018ilb}, and both for $d=1$ and $d=2$
the invariant connections are less constrained than in higher
dimensions. In this appendix we shortly go over the various cases and
point out the similarities and subtle differences with the generic
case discussed in the main text.

\subsection{$d=2$}

The case of two spatial dimensions is special, since it is the unique
dimension where there is a second rotationally invariant 2-tensor;
apart from the generic  $\delta_{ab}$ one now also has
$\epsilon_{ab}$. This extra tensor can appear in the symmetry algebra,
as well as in the invariant connection.

\subsubsection{$\calm_{(\a,\b)}$}

First we discuss the standard kinematical homogeneous spaces of
galilean type $\calm_{(\a,\b)}$, i.e. those based on the algebra
$\mathfrak{k}_{(\a,\b)}$, defined in
\eqref{isotropybrackets} and \eqref{galbrackets}. The homogeneous space
itself is constructed as in the higher dimensional cases in main text.
A subtle difference is in the classification of invariant affine
connections on this homogeneous space. A short calculation reveals
them to take the following form in modified exponential coordinates
\begin{equation}
  \begin{split}
    \Gamma^t_{tt} &=  \kappa + \iota  \\
    \Gamma^a_{tb} &= \delta^a_b \kappa + \kappa' \epsilon_{ba}\\
    \Gamma^a_{bt} &= \delta^a_b (\beta + \iota) - \kappa' \epsilon_{ba} \\
    \Gamma^a_{tt} &= \alpha x^a
  \end{split}
\end{equation}
Comparing to \eqref{eq:invcon} one sees that there is one
additional free parameter $\kappa'$. However, just like the other two
Nomizu parameters $\kappa$ and $\iota$ it drops out of the
autoparallel equation, which upon fixing the parameter as $\sigma=t$,
takes the form
\begin{align}
\ddot{x}^a + \beta \dot{x}^a + \alpha x^a = 0
\end{align}
This is exactly the same damped oscillation equation as in higher
dimensions and so the analysis of the symmetries and conserved charges
is identical to that of the main text.

\subsubsection{$\tilde \calm_{(\gamma,\chi)}$}

The existence of $\epsilon_{ab}$ leads to an additional class of
kinematical algebras of galilean type in two spatial dimensions. The
additional homogeneous kinematical spacetimes are called
S12$_{\gamma,\chi}$ in \cite{Figueroa-OFarrill:2018ilb}, but we will
refer to them as $\tilde \calm_{(\gamma,\chi)}$. Their underlying
kinematical Lie algebra $\tilde{\mathfrak{k}}_{(\gamma,\chi)}$ shares
the brackets \eqref{isotropybrackets} with
the generic algebra $\mathfrak{k}_{(\gamma,\chi)}$ , but the
non-vanishing brackets in \eqref{galbrackets} get replaced by
\begin{equation}
[H,B_a]=-P_a\qquad [H,P_a]=\gamma B_a+(1 + \gamma) P_a - \chi \epsilon_{ab} (P_b + B_b )\label{modgalbrackets}
\end{equation}
The associated homogeneous space is then similarly defined as $\tilde
\calm_{(\gamma,\chi)}=\tilde \calk_{(\gamma,\chi)}/\calh$. As in the
generic case one defines modified exponential coordinates
\eqref{modexp}. In these coordinates one can compute the vector fields
generating the $\tilde \calk_{(\gamma,\chi)}$ action:
\begin{equation}\label{eq:strvec}
  \begin{split}
    \xi_{J_{ab}} &= x^b \partial_a - x^a \partial b\\
    \xi_H &= \partial_t\\
    \xi_{B_a} &= A(t) \partial_a + B(t) \epsilon_{ab}  \partial_b\\
    \xi_{P_a} &= \dot{A}(t) \partial_a +\dot{B}(t) \epsilon_{ab} \partial_b,
  \end{split}
\end{equation}
where now
\begin{equation}
  \begin{split}
    A(t) &= \frac{e^{-t}(\gamma - 1)+ e^{-t\gamma} (\chi \sin(t\chi) - (\gamma -1)\cos(t\chi)) }{(\gamma - 1)^2 + \chi^2}\\
    B(t) &= \frac{e^{-t}\chi + e^{-t\gamma} (-\chi \cos(t\chi) - (\gamma -1)\sin(t\chi)) }{(\gamma - 1)^2 + \chi^2}.
  \end{split}
\end{equation}

The non-zero components of an invariant affine connection can then be
computed to be of the form
\begin{equation}
  \begin{split}
    \Gamma^t_{tt} &=  (\kappa+\iota) \\
    \Gamma^a_{tb} &=  (\kappa\delta^a_b + \kappa' \epsilon_{ba})\\
    \Gamma^a_{bt} &= (1+\gamma)\delta^a_b + \chi \epsilon_{ab} + (\iota \delta^a_b - \kappa' \epsilon_{ba})\\
    \Gamma^a_{tt} &= \gamma x^a + \chi \epsilon_{ab} x^b,
  \end{split}
\end{equation}
where $\kappa, \kappa', \iota$ are three unconstrained, real constants.

Upon fixing the parameter $\sigma=t$ the autoparallel equation associated to this invariant connection reads
\begin{align}
\ddot{x}^a + \dot{x}^a + \gamma(x^a + \dot{x}^a) + \chi \epsilon_{ab}(x^b + \dot{x}^b) &= 0\label{strangeq}
\end{align}
Note that these equations have a rather different\footnote{Formally
  the equations \eqref{strangeq} are some complexification of the
  equations \eqref{dampeq}, since by introducing $z=x^1+i x^2$ and
  $\alpha = \gamma - i \chi, \beta = 1 + \gamma - i \chi$ the
  equations \eqref{strangeq} take the form $\ddot z+\beta \dot
  z+\alpha z=0$. Indeed also the the algebra
  $\tilde{\mathfrak{k}}_{(\gamma,\chi)}$ can formally be brought into
  the form  $\mathfrak{k}_{(\a,\b)}$ via a complexification of boost
  and translation generators. I.e. defining $\mathbb{B} = B_1 - i
  B_2$, $\mathbb{P} = P_1 - i P_2$ one finds $[H, \mathbb{B}] = -
  \mathbb{P}$ and $ [H, \mathbb{P}] = \alpha\, \mathbb{B} +
  \beta\, \mathbb{P} $.} structure than that of the damped
harmonic oscillator \eqref{dampeq}.

Somewhat surprisingly there exists a one-parameter family of
Lagrangians (not related by a total derivative term) with
\eqref{strangeq} as their Euler--Lagrange equations:

\begin{equation}
L_{\theta}=\frac{e^{(1+\gamma)t}}{2}M_{ab}(\theta)(\dot x^a \dot x^b-A_{bc}x^ax^c)
\end{equation}
where
\begin{equation}
(M_{ab})=m\begin{pmatrix}
\cos(\chi t+\theta) & \sin(\chi t+\theta)\\
\sin(\chi t+\theta) & -\cos(\chi t+\theta)
\end{pmatrix}\quad (A_{ab})=\begin{pmatrix}
\gamma&\chi\\
-\chi&\gamma
\end{pmatrix} 
\end{equation}
One easily verifies that the actions $S_\theta=\int L_\theta dt$
are invariant under the transformations generated by $\xi_{J_{ab}}$,
$\xi_{B_a}$ and $\xi_{P_a}$ listed in \eqref{eq:strvec} and these thus
constitute symmetries in the usual sense. Under a time translation one
however finds
\begin{equation}
\delta_H S_\theta=(1+\gamma) S_\theta+\chi S_{\theta+\frac{\pi}{2}}\label{gensym}
\end{equation}
In the first term we recognize a homothetic rescaling of the action,
as we encountered in the main text. The second term mixes the action
$S_{\theta}$ with another action $S_{\theta+\frac{\pi}{2}}$. So time
translations are no longer a symmetry in the standard sense. But since
$S_{\theta}$ and $S_{\theta+\frac{\pi}{2}}$ both share the same
Euler--Lagrange equations, it follows that indeed the time-translations
leave these Euler--Lagrange equations invariant -- as can be directly
verified from the form \eqref{strangeq} -- and so \eqref{gensym} still
is a symmetry in a more general sense.

One could repeat the Hamiltonian analysis and try to find a way to
represent the symmetry algebra in terms of conserved charges. It is
rather straightforward to compute canonical momenta $\pc_a$ and their
transformation under the symmetries. One can then verify that indeed
the corresponding phase space space vector fields reproduce the Lie
algebra $\tilde{\mathfrak{k}}_{(\gamma,\chi)}$. What complicates a
further analysis however is that the  vector field corresponding to
time translations is no longer symplectic, and not even homothetic
symplectic. Indeed, it mixes the canonical symplectic form with a
non-canonical one. It would be interesting to understand this more
general notion of symplectic transformation, construct an analog of
hamiltonian vector fields and a corresponding generalized notion of
Poisson bracket. I.e. extending the discussion of Appendix~\ref{sec:homap}
to these more general transformations. This would however take us too
far from the main topic of this paper and so we leave this as an
interesting open problem.

\subsection{$d=1$}

The case of one spatial dimension is special, since there are no
rotations in this case. This does not change anything from the point
of view of the kinematical algebras or homogeneous spaces, which
remain only of the type $\calm_{(\a,\b)}$. In particular the vector
fields generating the group action remain the same, i.e.,
\eqref{kilvec}. A small but not completely trivial change is the
presence of an extra freedom in the invariant connections, which take
the form
\begin{equation}
  \begin{split}
    \Gamma^t_{tt} &= (\kappa + \iota)  \\
    \Gamma^1_{t1} &=  \kappa \\
    \Gamma^1_{1t} &= (\beta + \iota)\\
    \Gamma^1_{tt} &= \alpha x - \psi.
  \end{split}
\end{equation}
Comparing to the case in generic dimensions, \eqref{eq:invcon} one sees there is an additional free parameter $\psi$.

Unlike in all other cases, the Nomizu parameter $\psi$, particular to $d=1$, does not drop out of the autoparallel equation:
\begin{align}
\ddot{x} + \beta \dot{x} + \alpha x = \psi\label{pseq}
\end{align}
The extra Nomizu parameter corresponds physically speaking to an extra constant force
$m\psi$.

We should point out that when $\alpha\neq 0$ adding $\psi$ is rather trivial, since one can obtain all expressions from the $\psi=0$ case simply by the replacement $x\rightarrow  x-\alpha^{-1}\psi$. This is true for the action \eqref{BCK}, and the conserved charges (\ref{ch1}-\ref{ch2}), which thus take the form
\begin{equation}\label{eq:dPdB}
  \begin{split}
    \cP&=\dot F \pc-me^{\beta t}\ddot F (x-\alpha^{-1}\psi)\\
    \cB&= F \pc-me^{\beta t}\dot F (x-\alpha^{-1}\psi).
  \end{split}
\end{equation}
Also the Euler vector field \eqref{explE} gets shifted,
\begin{equation}
\tilde E=\frac{1}{2}(x-\alpha^{-1}\psi)\partial_x + \frac{p}{2}\partial_p\,
\end{equation}
and one should consider the canonical Hamiltonian
\begin{equation}
\tilde \hc=\hc+m \frac{\psi^2}{2\a}e^{\beta t}=\frac{e^{-\beta t}}{2m}\pc^2+\frac{m\a e^{\beta t}}{2}( x-\alpha^{-1}\psi)^2\,.
\end{equation}
This to guarantee one has the crucial relationship $\call_{\tilde
	E}\tilde \hc=\tilde \hc$, see  Appendix~\ref{sec:homap}. One additionally verifies that the vector
field generating the phase space time translation takes the form
\begin{equation}
\Xi_H=X_{(\beta,\cH)}=\beta \tilde E+X_\cH\qquad \cH=-(\tilde \hc+\frac{\beta}{2}(x-\a^{-1}\psi)\pc)
\end{equation}
The realization of the kinematical
algebra $\mathfrak{\calk}_{(\a,\b)}$ in terms of conserved charges
then goes through as in the main text.

Somewhat to our surprise, the case $\alpha=0$ is subtly different. One
still has an action for \eqref{pseq}, which in this case reads
\begin{align} S = \int \frac{m e^{\beta t}}{2} ( \dot{x}^2 + 2\psi
  x)\, dt\label{1dlag} \end{align} For the spatial translation and
boost we find satisfying and rather standard results. There exist the
conserved charges\footnote{Note that both expressions in \eqref{eq:dPdB}
  are singular in the $\alpha\rightarrow0$ limit. This singularity can
  be removed by adding to the expressions for $\cP$ and $\cB$ the
  constants $\alpha^{-1}\beta m\psi$ and $-\alpha^{-1} m\psi$,
  respectively. The charges in \eqref{p1b1} are then the
  $\alpha\rightarrow 0$ limit of these 'regularized' charges.}
\begin{eqnarray}\label{p1b1}
  \begin{split}
    \cP &= e^{-\beta t} \pc + m (\beta x - \psi t)\\
    \cB &= \beta^{-1}(1-e^{-\beta t})\pc - m x +m\psi \beta^{-2}(1+\beta t -e^{\beta t})
  \end{split}
\end{eqnarray}
and the phase space transformations are generated by the corresponding
hamiltonian vector fields $X_\cP$ and $X_\cB$ respectively.
But when $\beta\neq 0$, there does not exist an Euler
field and charge $\cH$ for which $\Xi_H=\beta E+X_\cH$ while at the same time also the
conditions \eqref{symconds} are met. This implies we cannot represent
the algebra $\mathfrak{k}_{(0,\beta)}$ in terms of conserved charges
when $\beta\neq 0$ and $\psi\neq 0$.

The special case when both $\alpha=\beta=0$ does have a (simple)
solution. In this case the time translation vector field is
hamiltonian: $\Xi_H=X_\cH$ with
\begin{equation}
  \cH=-\hc=-\frac{\pc^2}{2m}+m\psi x 
\end{equation}
Since $\hc$ is time independent this charge is indeed conserved. In
addition one can set $\beta=0$ in
\eqref{p1b1} to get conserved charges
\begin{equation}
  \begin{split}
    \left.\cP\right|_{\alpha=\beta=0} &= \pc - \psi m t\\
    \left.\cB\right|_{\alpha=\beta=0} &= \pc t - mx -\psi m\frac{t^2}{2}.
  \end{split}
\end{equation}
We thus uncover the curiosity that in one dimension a free particle,
i.e., $\alpha=\beta=0$, remains Galilei-invariant even in the presence
of a constant force $m\psi$. In other words, the fact that in $d>1$
the presence of a constant force breaks this invariance is only due to
the fact that it breaks rotational invariance, which is not an issue
when $d=1$. Note however that the Bargmann central extension appears slightly differently when $\psi\neq 0$:
	\begin{equation}
	\{P,B\}=m\qquad \{H,B\}=-P\qquad \{H,P\}=\psi m
	\end{equation}
	It can be put into a standard form by defining $\tilde H=H+\psi B=-\frac{(p-m\psi t)^2}{2}$:
	\begin{equation}
	\{P,B\}=m\qquad \{\tilde H,B\}=-P\qquad \{\tilde H,P\}=0
	\end{equation}
	Actually this redefinition of time translations is an isomorphism of the 1d Galilei algebra $\mathfrak{k}_{(0,0)}$.
	
\section{Conformally equivalent Eisenhart lifts}\label{confapp}

In Section \ref{sec:NCbarg} we reviewed how mechanical motion can be
described in a covariant fashion using Newton-Cartan geometry and how
that description in turn is equivalent to null geodesic motion with
respect to a particular Lorentzian metric. In this appendix we shortly
recall that these descriptions are unique only up to a conformal
redefinition. We refer to \cite{Bekaert:2013fta} for a complete
discussion, here we simply mention how some key formulae of Section
\ref{sec:NCbarg} behave under such a conformal redefinition and then
focus on the case of interest in this paper, namely the damped
harmonic motion describing a free particle on $\calm_{(\a,\b)}$, and
how some results in this paper are related to those in
\cite{Figueroa-OFarrill:2022ryd}.

The starting observation is that the Type I NC geometries $(\tau_\mu,
h^{\m\n}; \hat \tau^\mu, \bar h_{\m\n})$ and $(\tau_\mu^{\mathrm{c}},
h^{\m\n}_{\mathrm{c}}; \hat \tau_{\mathrm{c}}^\mu, \bar
h_{\m\n}^{\mathrm{c}})$ lead to the same action \eqref{NCaction} if
they are conformally related as
\begin{equation}
  \tau_\mu^{\mathrm{c}}=e^\psi\tau_\mu\quad h^{\mu\nu}_{\mathrm{c}}=e^{-\psi}h^{\mu\nu}\quad \hat\tau^\mu_{\mathrm{c}}=e^{-\psi}\hat\tau^\mu\quad \bar h_{\mu\nu}^{\mathrm{c}}=e^{\psi}\bar h_{\mu\nu} 
\end{equation}
for an arbitrary function $\psi$. So really the particle action
\eqref{NCaction} only depends on a conformal class of type I NC
geometries.

Under such a conformal redefinition of the NC geometry essentially all results of Section
\ref{sec:NCbarg} remain valid, but some of the parameters will depend
on the choice of conformal prefactor. For example, the parameters characterizing the
invariance of the action under symmetries, as in
\eqref{NCtransformation}, transform as
\begin{equation}
\zeta_\mathrm{c}=\zeta+\call_\xi\psi\qquad \lambda_\mathrm{c}=\lambda\qquad K_\mathrm{c}=K\label{redefzeta}
\end{equation}

Since the lift \eqref{hatxi} of the vector fields generating the
symmetries only depends on $\lambda$ and $K$ it remains invariant
under a conformal redefinition. In turn this implies that the lifted
symmetry algebra is independent of the choice of conformal prefactor
$\psi$. Of course this is not the case for the Lorentzian metric
\eqref{nullred}: unsurprisingly, it transforms by a conformal
rescaling
\begin{equation}
g_{AB}^\mathrm{c}=e^\psi g_{AB}.\label{redefmet}
\end{equation}
Note that since $\psi$ is, by construction, independent of $u$, the
vector field $k=\partial_u$ remains null and Killing for all choices
of conformal factor.

The transformations \eqref{redefzeta} and \eqref{redefmet} are
compatible in that they leave \eqref{gtransfo} invariant, i.e., one
also has that
\begin{equation}
  \call_{\hat \xi}g_{AB}^\mathrm{c}=(\lambda_c+\zeta_c) g_{AB}^\mathrm{c},\label{confkil}
\end{equation}
so that for any choice of conformal factor the lifted symmetries will
be conformal Killing vectors.   Let us remark that since $\zeta$ changes
under a change of conformal factor, see \eqref{redefzeta}, one may be
able to find special conformal factors for which all
$\lambda_\mathrm{c}+\zeta_{\mathrm{c}}$ are constant and the lifted
symmetries are homotheties, or even such that
$\lambda_\mathrm{c}+\zeta_{\mathrm{c}}=0$ for all vector fields so
that the lifted symmetries become isometries.  This turns out to be
possible for the geometries of interest.

Let us now specialize to motion on $\calm_{(\a,\b)}$, the case of
interest in this paper. Allowing for a generic conformal factor, the
NC geometry (\ref{betaNC}, \ref{betatriple}) becomes
\begin{equation}
  \begin{split}
    \tau_\mu^\mathrm{c} &= e^\psi\delta^t_\mu\\
    h_\mathrm{c}^{\mu\nu} &= e^{-\beta t-\psi}\delta_a^\mu \delta_a^\nu\\
    \hat\tau^\mu_\mathrm{c} &= e^{-\psi}\delta_t^\mu\\
    \bar h_{\mu\nu}^\mathrm{c} &= e^{\beta t+\psi}\delta^a_\mu
    \delta^a_\nu-\a e^{\beta t+\psi} x^a x^a\delta^t_\mu\delta^t_\nu\,.
  \end{split}
\end{equation}
The Lorentzian metric describing the Eisenhart lift \eqref{bargstr} then gains an overall conformal prefactor:
\begin{equation}
  ds^2_\mathrm{c}=e^\psi\left(-2du dt-\a e^{\beta t}x^ax^adt^2+e^{\beta t}dx^a dx^a\right)\qquad k=\partial_u\label{appmet}
\end{equation}
To make the equivalence to \cite{Figueroa-OFarrill:2022ryd} explicit,
make the coordinate transformation
\begin{equation}
  u'=u+\Lambda\qquad \Lambda=\frac{\beta}{2} e^{\beta t}x^a x^a\label{coordtransfo}
\end{equation}
Note that this is a special type of coordinate transformation
corresponding to the redefinitions of the NC doublet, as mentioned in
\eqref{doubletredef}. The Bargmann structure \eqref{appmet} then
becomes
\begin{equation}
  ds^2_\mathrm{c}=e^\psi\left(-(2du'+\a e^{\beta t}x^ax^adt)dt+e^{\beta t}(dx^a+\beta x^a dt)(dx^a+\beta x^a dt)\right)\qquad k=\partial_{u'}\label{bargstr2}
\end{equation}
Upon the choice $\psi=-\beta t$ this becomes exactly the metric
described in Appendix A of \cite{Figueroa-OFarrill:2022ryd}.  In
conclusion, while the lifted vector fields \eqref{liftkilvec}, that
form the algebra \eqref{lalg}, generate homotheties of the metric
\eqref{bargstr}, these same transformations are isometries of
\eqref{bargstr2} when $\psi=-\beta t$. This follows from
\eqref{confkil} and the fact that
$\zeta_{H}^\mathrm{\,c}=-\beta=-\lambda_{H}^\mathrm{c}$, while
$\zeta_{H}=0\,, \lambda_H=\beta$.

\bibliographystyle{utphys}
\bibliography{kinalg}
\end{document}